\documentclass[twocolumn,hidelinks]{article}

\usepackage{amsmath,amssymb}
\usepackage{lineno}
\modulolinenumbers[1]
\usepackage{tikz}
\usetikzlibrary{arrows}
\usepackage{pgfplotstable}
\pgfplotsset{compat=1.14}
\usepackage{booktabs}
\usepackage{xfrac}
\usepackage{siunitx}
\usepackage{enumitem}

\def\intdomd{\Omega^\delta}
\def\extdomd{\widehat{\Omega}^\delta}
\def\uelast{\underline{u}}
\def\uperid{u}

\newcommand\corr[1]{{\color{black}{#1}}} 
\newcommand\edit[1]{{\color{black}{#1}}}

\begin{document}



\title{On the treatment of boundary conditions for bond-based \corr{peridynamic} models}

\author{Serge Prudhomme \\ Department of Mathematics and Industrial Engineering, Polytechnique Montr\'eal, \\
C.P. 6079, succ.\ Centre-ville, Montr\'eal, Qu\'ebec H3C~3A7, Canada \\ serge.prudhomme@polymtl.ca \\\\
Patrick Diehl\\Center for Computation and Technology, Louisiana State University, \\
Digital Media Center, 340 E.\ Parker Blvd, Baton Rouge, LA 70803, USA\\http://orcid.org/0000-0003-3922-8419}
\date{}




\maketitle

\begin{abstract}
In this paper, we propose two approaches to apply boundary conditions for bond-based \corr{peridynamic} models. There has been in recent years a renewed interest in the class of so-called non-local models\corr{, which include peridynamic models}, for the simulation of structural mechanics problems as an alternative approach to classical local continuum models. However, a major issue, which is often disregarded when dealing with this class of models, is concerned with the manner by which boundary conditions should be prescribed. Our point of view here is that classical boundary conditions, since applied on surfaces of solid bodies, are naturally associated with local models. The paper describes two methods to incorporate classical Dirichlet and Neumann boundary conditions into bond-based peridynamics. The first method consists in artificially extending the domain with a thin boundary layer over which the displacement field is required to behave as an odd function with respect to the boundary points. The second method resorts to the idea that \corr{peridynamic} models and local models should be compatible in the limit that the so-called horizon vanishes. The approach consists then in decreasing the horizon from a constant value in the interior of the domain to zero at the boundary so that one can directly apply the classical boundary conditions. We present the continuous and discrete formulations of the two methods and assess their performance on several numerical experiments dealing with the simulation of a one-dimensional bar.  
\end{abstract}




\section{Introduction}

In recent years, novel non-local modeling theories have been developed for the simulation of problems in structural and fracture mechanics. \corr{Peridynamic theory~\cite{Silling-JMPS-2000}}, which can be viewed as a generalization of classical continuum mechanics, provides for instance such a class of non-local models. \corr{Peridynamics} does not involve differentials of displacement fields, which makes it an attractive framework for the modeling and simulation of fracture mechanics applications. Validation of the peridynamic theory has been undertaken using experimental data from a variety of wave propagation applications as well as crack initiation and propagation experiments, see e.g.~\cite{Diehl2019}. However, major issues identified in that review include the so-called surface or skin effect~\cite{bobaru2009-1D,sarego2016linearized,le2018surface} and the manner by which boundary conditions should be modeled within a non-local model. \corr{This was recognized early on during the development of peridynamics as Silling and Askari in~\cite{Silling-Askari-CS-2005}, for instance, explained that displacement and traction boundary conditions should be prescribed within a layer of finite thickness under the surface of a solid body. Unfortunately, such boundary data cannot be assumed to be known in practical applications of interest. Several methods have since been proposed, see e.g.~\cite{du2016nonlocal,aksoylu2019nonlocal,gu2018revisit,madenci2018state,madenci2018weak,gunzburger2010nonlocal,aksoylu2010results}, in order to circumvent the lack of data and correct the spurious surface effects that are created as a result in the vicinity of the boundaries. Starting from the fact that data is only available at the surface of a solid body, we believe that boundary conditions for peridynamic models should be prescribed in a manner similar to that for local classical continuum models.} We \corr{study} in this paper two approaches to incorporate classical Dirichlet and Neumann boundary conditions into a one-dimensional \corr{linearized bond-based peridynamic model~\cite{Silling-JMPS-2000}}. \corr{Our main objective in this paper is to mathematically analyze the resulting models, and more specifically, to verify that their solutions converge to that of the compatible linear elasticity model with respect to the so-called peridynamic horizon and discretization parameters.}  

The first approach, referred here to as the extended domain method (EDM), relies on the idea to artificially extend the domain by a thin layer of the size of peridynamic horizon. \corr{The idea is not new and is similar to that proposed by the authors in~\cite{aksoylu2011variational,seleson2013interface,seleson2016convergence, gerstle2005peridynamic,madenci2014peridynamic,oterkus2014peridynamic}, among others. In~\cite{aksoylu2011variational,seleson2013interface,seleson2016convergence} though, the thin layer was introduced within the solid body and subjected to Dirichlet-like non-local boundary conditions, that is, the displacement field in the layer was either arbitrarily set to zero or was constrained to match the values of the exact solution in the case of a manufactured problem. Zhou and Du~\cite{Zhou-Du-SIAMJNA-2010} proposed to constrain the solutions to be either odd or even from either side of the boundary, in order to mimic homogeneous Dirichlet or Neumann boundary conditions, respectively. We propose here to constrain the displacement field in the artificial layer to satisfy odd extensions with respect to the boundary points so that one can handle both Dirichlet and Neumann non-homogeneous boundary conditions.}

The second approach, referred here to as the variable horizon method (VHM), proposes to vary the peridynamic horizon from a constant value in the interior of the domain to zero as one approaches the boundary. \corr{The approach was considered in~\cite{bobaru2009-1D,bobaru2011adaptive} for the treatment of boundary conditions along with adaptive mesh refinement.} The method is motivated by the fact that the local model and non-local model are consistent and compatible in the limit that the horizon vanishes, so that one can justify the use of classical Dirichlet and Neumann boundary conditions to be prescribed at the boundary of the domain. \corr{Varying the horizon size necessitates however a scaling of the material parameter for the peridynamic model. We propose here an approach similar to~\cite{bobaru2009-1D,bobaru2011adaptive} but with a scaling in 1D that appears to be different from that suggested in~\cite{bobaru2009-1D}. We also mention that the variable horizon approach was utilized in~\cite{seleson2013interface,silling2015variable,silling15Variablehorizonperidynamicmedium} for the treatment of interface problems in which different peridynamic models or a continuum model and a peridynamic model are coupled together.}

Performance and comparison of the two methods are investigated on several numerical examples involving a one-dimensional linear elastic bar. In particular, we carry out a $\delta$-convergence analysis~\cite{bobaru2009-1D,seleson2016convergence} by considering four test cases in which the manufactured solutions are linear, quadratic, cubic, and quartic, in order to put in evidence the differences and similarities between the methods. The EDM and VHM solutions are also compared with the solution of the local linear elasticity model (LLEM). All models are discretized here using the finite difference method in order to allow for consistent comparison and are assessed using the maximum relative error with respect to the exact solution of the linear elastic local problem. In all cases, we verify that the numerical rates of convergence at which the peridynamic solutions converge to the linear elasticity solution are consistent with the theoretical rates. We also present additional examples in order to study the effect of a correcting factor for the extended domain method, to carry out a $m$-convergence analysis, and to consider the case of a solution with a very steep gradient in the vicinity of one boundary point.

The paper is organized as follows. Following the introduction, we present in Section~\ref{Sect:modelproblem} the model problem based on the classical linear elasticity theory and corresponding linearized bond-based \corr{peridynamic} model for the simulation of a one-dimensional bar held fixed at one end and subjected to a traction at the other end. We proceed in Section~\ref{Sect:correctedfoces} with the description of well-known spurious effects near boundaries when using non-local models and assess a correction method to alleviate the issue. We introduce in Section~\ref{Sect:extendeddomain} and Section~\ref{Sect:variable-horizon} the continuous and discrete formulations of the extended domain method and the variable horizon method, respectively, and assess their performance in Section~\ref{Sect:numericalexamples} on several one-dimensional numerical examples. We complete the paper in Section~\ref{Sect:conclusion} with concluding remarks and perspectives.

\section{Model problem and preliminaries}
\label{Sect:modelproblem}

The model problem \corr{consists in finding the static equilibrium of} a bar, of unit length and constant cross-sectional area, held fixed at one end and subjected to a longitudinal traction at the other \corr{end}. We shall suppose here that the deformations in the bar are infinitesimally small and can be adequately described by the theory of linear elasticity.  We present below the local model based on classical elasticity theory and the non-local model counterpart built upon the linearized bond-based \corr{peridynamic} theory~\cite{Silling-JMPS-2000}.

\subsection{Continuum local model}

Let $\Omega = (0,1)$. The continuum local problem consists in finding the displacement $\uelast$ in the bar such that:
\begin{align}
\label{eq:1dlinearelasticity}
- EA \uelast'' = f_b, &\quad \forall x \in \Omega, \\
\label{eq:Dirichlet}
\uelast = 0, &\quad \text{at}\ x=0,\\
\label{eq:Neumann}
EA\uelast' = g, &\quad \text{at}\ x=1,
\end{align}
where \corr{$E$ and $A$ are the constant modulus of elasticity and cross-sectional area of the bar, respectively, $g \in \mathbb R$ is the traction force applied at end point $x=1$, and the scalar function $f_b=f_b(x)$ is the external body force density (per unit length).} We suppose here that $f_b$ is chosen sufficiently smooth so that regularity in the solution is not an issue when comparing the solutions from the two models. \corr{Note that we use the notation $\uelast$ for the solution to the classical elasticity problem in order to emphasize that it may be different from the peridynamic solution $\uperid$ introduced below.}

\subsection{Peridynamic model}

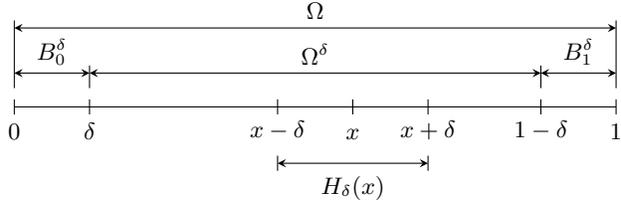
\begin{figure}[tbp]
\centering
\small
\begin{tikzpicture}
\draw (0,0) -- (8.0,0);
\draw (0,-0.1) -- (0,0.1);
\draw (1,-0.1) -- (1,0.1);
\draw (3.5,-0.1) -- (3.5,0.1);
\draw (4.5,-0.1) -- (4.5,0.1);
\draw (5.5,-0.1) -- (5.5,0.1);
\draw (7.0,-0.1) -- (7.0,0.1);
\draw (8.0,-0.1) -- (8.0,0.1);
\node[above] at (0.0,-0.55) {$0$};
\node[above] at (1.0,-0.55) {$\delta$};
\node[above] at (3.5,-0.55) {$x-\delta$};
\node[above] at (4.5,-0.55) {$x$};
\node[above] at (5.5,-0.55) {$x+\delta$};
\node[above] at (7.0,-0.55) {$1-\delta$};
\node[above] at (8.0,-0.55) {$1$};
\draw[arrows=<->, >=stealth]  (0.0,1.05) -- (8.0,1.05);
\draw  (0.0,0.25) -- (0.0,1.15);
\draw  (8.0,0.25) -- (8.0,1.15);
\node[above] at (4.00,1.05) {$\Omega$};
\draw[arrows=<->, >=stealth]  (0.0,0.45) -- (1.0,0.45);
\draw[arrows=<->, >=stealth]  (1.0,0.45) -- (7.0,0.45);
\draw[arrows=<->, >=stealth]  (7.0,0.45) -- (8.0,0.45);
\draw  (1.0,0.25) -- (1.0,0.55);
\draw  (7.0,0.25) -- (7.0,0.55);
\node[above] at (4.00,0.45) {$\intdomd$};
\draw[arrows=<->, >=stealth]  (3.5,-0.75) -- (5.5,-0.75);
\draw  (3.5,-0.65) -- (3.5,-0.85);
\draw  (5.5,-0.65) -- (5.5,-0.85);
\node[below] at (4.5,-0.8) {$H_\delta(x)$};
\node[above] at (0.5,0.45) {$B_0^{\delta}$};
\node[above] at (7.5,0.45) {$B_1^{\delta}$};
\end{tikzpicture}
\caption{Definition of the computational domains for the \corr{peridynamic} model.}
\label{Fig:peridynamicsdomains}
\end{figure}

Let $\delta > 0$ denote the so-called horizon of the \corr{peridynamic} model and let $H_\delta(x) = (x-\delta, x+\delta)$ be the subdomain of the neighboring particles within the horizon. We observe that for any given point $x$ in the interval $\intdomd = (\delta,1-\delta)$, we have that $H_\delta(x) \subset \Omega$. We also introduce the boundary domains $B_0^\delta = (0,\delta)$ and $B_1^\delta = (1-\delta,1)$ such that $\overline{\Omega} = \overline{B_0^\delta \cup \intdomd \cup B_1^\delta}$, as show in Figure~\ref{Fig:peridynamicsdomains}. The general formulation, \corr{in one or higher dimension,} of the \corr{linearized microelastic  model~\cite{Silling-JMPS-2000}} is given by:
\begin{equation}
\label{eq:1dperidynamics}
\int_{H_\delta(x) \cap \Omega} \kappa \frac{\xi \otimes \xi}{\| \xi \|^3} (\uperid(y) - \uperid(x)) dy + f_b(x) = 0, 
\end{equation}
where $\kappa$ is the parameter that characterizes the stiffness of the ``bonds'' between point $x$ and the neighboring points $y \in H_\delta(x)$, $\xi$ is the vector between two material points in the reference configuration, i.e.\ $\xi = y-x$, $\| \xi \|$ is the Euclidean norm of vector $\xi$, and $\uperid(x)$ is the displacement of $x$ in the deformed configuration. In the case of the one-dimensional bar, the above integral at a point $x\in \intdomd$ can be rewritten as:
\begin{equation}
\label{eq:1dperidynamics2}
\int_{x-\delta}^{x+\delta} \kappa \frac{\uperid(y) - \uperid(x)}{|y-x|} dy + f_b(x) = 0.
\end{equation}

Ideally, the material parameter $\kappa$ could be identified from experimental data. However, one may be interested in the value of $\kappa$ such that the solution of the linearized microelastic brittle \corr{peridynamic} model is the same as that of the continuum local model. One may do so by matching the strain energy of the two models~\cite{Silling-JMPS-2000}. Alternatively, one can try to recover equation~\eqref{eq:1dlinearelasticity} from~\eqref{eq:1dperidynamics} when taking the limit $\delta \rightarrow 0$. Supposing that the displacement $\uperid(x)$ is sufficiently smooth, one can write, for all $y \in \intdomd$:
\begin{equation}
\begin{aligned}
\uperid(y) - \uperid(x) 
& = \uperid'(x) (y-x) + \frac{1}{2} \uperid''(x) (y-x)^2\\
& + \frac{1}{6} \uperid'''(x) (y-x)^3 + \mathcal O (|y-x|^4),
\end{aligned}
\end{equation}
so that the integral in~\eqref{eq:1dperidynamics} reduces to:
\begin{equation}
\label{eq:forceperidynamics}
\int_{x-\delta}^{x+\delta} \kappa \frac{\uperid(y) - \uperid(x)}{|y-x|} dy = \frac{\kappa\delta^2}{2} (\uperid''(x) + \mathcal O(\delta^2)).
\end{equation}
Substituting the integral in~\eqref{eq:1dperidynamics} with \corr{the} above result yields:
\begin{equation}
\label{eq:expansion}
- \frac{\kappa\delta^2}{2} (\uperid'' + \mathcal O(\delta^2)) = f_b, \qquad \forall x \in \intdomd.
\end{equation}
By taking the limit when $\delta \rightarrow 0$, one then recovers the differential equation~\eqref{eq:1dlinearelasticity} pointwise whenever $\kappa$ is chosen as:
\begin{equation}
\label{eq:kappa}
\frac{\kappa\delta^2}{2} = EA, \quad \text{that is}\
\kappa = \frac{2EA}{\delta^2},
\end{equation}
in agreement with~\cite{Seleson-Du-Parks-CMAME-2016}. In other words, the \corr{peridynamic} model can be used to describe the behavior in the bar as one is able to identify the value of parameter $\kappa$ that makes it compatible to the linear elasticity model. The main and well known issue with peridynamics is whether the model can also be employed in the region $\Omega \backslash \intdomd$, i.e.\ in the region of size $\delta$ along the boundary. A solution to this issue could be to transpose the Dirichlet condition~\eqref{eq:Dirichlet} and Neumann condition~\eqref{eq:Neumann}, prescribed at the boundaries of the bar, into volumetric boundary conditions to constrain the material in $B_0^\delta$ and $B_1^\delta$. However, it is still unclear how this can be achieved, and if it were, how to identify the corresponding boundary data. The point of view that we will follow here is that the notion of boundary conditions, which are prescribed on the surface of solid bodies, is naturally associated with local models. Our goal in the following section is to present some approaches to adapt the Dirichlet and Neumann conditions to the \corr{peridynamic} problem.  

\section{Corrected forces near free surfaces}
\label{Sect:correctedfoces}

In this section, we study the integral in~\eqref{eq:1dperidynamics} at a point in $B_0^\delta$, i.e.\ near the left boundary of the bar. The analysis is similar for a point in $B_1^\delta$ near the other boundary. Let $x \in B_0^\delta$. Since $H_\delta(x)$ extends beyond $\Omega$, the integral becomes:
\begin{equation}
\label{eq:forceboundary}
\begin{aligned}
& \int_{0}^{x+\delta} \kappa \frac{\uperid(y) - \uperid(x)}{|y-x|} dy\\
& = \int_{-x}^{\delta} \kappa \Big( \uperid'(x) \frac{\xi}{|\xi|} 
+ \frac{\uperid''(x)}{2}  \frac{\xi^2}{|\xi|} + \frac{\uperid'''(x)}{6}  \frac{\xi^3}{|\xi|}
+ \ldots \Big) d\xi\\
& = \kappa \uperid'(x) \int_{-x}^{\delta} \frac{\xi}{|\xi|} d\xi 
+ \frac{\kappa}{2} \uperid''(x) \int_{-x}^{\delta} |\xi| d\xi \\
&\qquad
+ \frac{\kappa}{6} \uperid'''(x) \int_{-x}^{\delta} |\xi| \xi d\xi
+ \ldots \\
& = \kappa (\delta-x) \uperid'(x) 
+ \frac{\kappa}{2} \frac{ \delta^2 + x^2}{2} \uperid''(x) \\
&\qquad 
+ \frac{\kappa}{6} \frac{ \delta^3 - x^3}{3} \uperid'''(x) 
+ \ldots
\end{aligned}
\end{equation}
where we have used the change of variable $\xi = y-x$.
When comparing \corr{the} above result to~\eqref{eq:forceperidynamics}, it is clear that truncation of the set $H_\delta(x)$ close to the boundary induces forces involving the first derivative of $\uperid$. \edit{This term vanishes in the case of free surfaces since zero traction, i.e.\ $EA\uperid' = 0$, is prescribed on the boundary. Moreover, the coefficient of $\uperid''$ is different from that in~\eqref{eq:forceperidynamics}. Since $0 \leq x \leq \delta$, the coefficient is actually smaller than for points in the interior of the domain, meaning that the material becomes softer as one approaches the free boundary, which is referred to as the so-called skin effect~\cite{bobaru2009-1D}}. 

\edit{Several methods to correct the skin effect, such as the volume correction method~\cite{Parks-Lehoucq-CPC-2008,silling2016introduction}, the force density method~\cite{oterkus2010peridynamic, madenci2014peridynamic}, the energy  method~\cite{oterkus2010peridynamic, madenci2014peridynamic}, the force normalization method~\cite{Macek-Silling-FEAD-2007}, or the position-aware method~\cite{mitchell2015position} have been compared in~\cite{le2018surface}. We just note that some of the methods seem somewhat heuristic and that the force normalization approach of 
Macek and Silling~\cite{Macek-Silling-FEAD-2007} appears to be valid only for two- and three-dimensional peridynamic models, as it involves an integral that is singular in the case of one-dimensional problems}. 

\edit{Looking at~\eqref{eq:forceboundary}, a simple approach would be to replace the parameter $\kappa$ by a position-dependent parameter $\bar\kappa(x)$ such that the coefficient of the second derivative matches that obtained for a point inside $\intdomd$, see~\eqref{eq:forceperidynamics}, or equivalently, allows one to recover the linear elasticity model, that is:}
\[
\frac{\bar\kappa(x)}{2} \frac{ \delta^2 + x^2}{2} = \frac{\kappa \delta^2}{2} = EA,
\]
which leads to:
\begin{equation}
\label{eq:correction}
\bar\kappa(x) = \kappa \frac{2 \delta^2}{\delta^2 + x^2}.
\end{equation}
\edit{It is notable that this result is equivalent to the correction factor that one obtains by the volume correction method. In fact, one would recover the same result by the energy method, at least in 1D.} Indeed, the method consists in identifying the parameter $\bar\kappa(x)$ such that the strain energy computed at a point $x$ near the boundary matches that computed at a point inside $\intdomd$ for a given deformation. The strain energy per unit length of a ``bond'' is given by:
\[
\omega(\corr{u}) = \frac{\bar\kappa(x)}{2} \frac{(\uperid(y) - \uperid(x))^2}{|y-x|}.
\]
It follows that the strain energy at a point $x \in B_0^\delta$ is provided by:
\[
\begin{aligned}
\int_0^{x+\delta} \frac{\bar\kappa(x)}{2}  \frac{(\uperid(y) - \uperid(x))^2}{|y-x|} dy
& \approx \frac{\bar\kappa(x)}{2} [\uperid'(x)]^2 \int_{-x}^{\delta} |\xi| d\xi \\
& \approx \frac{\bar\kappa(x)}{4} [\uperid'(x)]^2 ( \delta^2 + x^2).
\end{aligned}
\]
For a point inside the domain, i.e.\ $x \in \intdomd$, the strain energy reads:
\[
\begin{aligned}
\int_{x-\delta}^{x+\delta} \frac{\kappa}{2}  \frac{(\uperid(y) - \uperid(x))^2}{|y-x|} dy
& \approx \frac{\kappa}{2} [\uperid'(x)]^2 \int_{-\delta}^{\delta} |\xi| d\xi \\
& \approx \frac{\kappa}{2} [\uperid'(x)]^2 \delta^2.
\end{aligned}
\]
Matching those two energies for a given deformation $\uperid'$ naturally leads to~\eqref{eq:correction}.

\edit{Replacing $\kappa$ by $\bar{\kappa}(x)$ in~\eqref{eq:forceboundary} and using~\eqref{eq:correction}, one obtains the new expression of the integral:
\begin{equation}
\label{eq:forceboundarycorrected}
\begin{aligned}
& \int_{0}^{x+\delta} \kappa \frac{\uperid(y) - \uperid(x)}{|y-x|} dy\\
& \quad = \kappa \frac{2\delta^2(\delta-x)}{\delta^2+x^2} \uperid'(x)  
+ \frac{\kappa \delta^2}{2} \uperid''(x) \\
&\qquad\qquad 
+ \frac{\kappa}{9} \frac{ \delta^2 (\delta^3 - x^3)}{\delta^2+x^2} \uperid'''(x)  
+ \ldots \\
&\quad =
\frac{\kappa \delta^2}{2} \bigg( 
\frac{1-(x/\delta)}{1+(x/\delta)^2} \frac{4}{\delta} \uperid'(x) + \uperid''(x) \\
&\qquad\qquad
+ \frac{1-(x/\delta)^3}{1+(x/\delta)^2} \frac{2\delta }{9} \uperid'''(x) + \ldots
\bigg)
\end{aligned}
\end{equation}
Therefore, if $\uperid'(x) = 0$, such as near free surfaces, the peridynamic model leads to:
\begin{equation}
\label{eq:expansionboundary}
- \frac{\kappa\delta^2}{2} (\uperid'' + \mathcal O(\delta)) = f_b, \qquad \forall x \in B_0^\delta.
\end{equation}
In other words, due to the loss of symmetry of the integration domain, the peridynamic model provides an approximation of the classical linear elasticity theory of order one in $\delta$ in the vicinity of free surfaces, to be compared to $\mathcal O(\delta^2)$ in the interior of the domain, see~\eqref{eq:expansion}.}

Nevertheless, we emphasize here that the above correction approach is still unsuitable for problems involving Dirichlet boundary conditions or non-homogeneous Neumann boundary conditions, since the first derivative of the solution does not necessarily vanish near the boundaries in these cases. We present in the next two sections alternative approaches to alleviate boundary effects, namely the extended domain method, if the horizon must be kept constant everywhere, and the variable horizon method, if the size of the horizon is allowed to vary in the domain.  

\section{Extended domain method}
\label{Sect:extendeddomain}

\subsection{Continuous formulation}

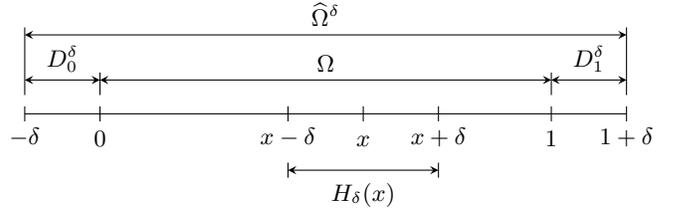
\begin{figure}[tbp]
\centering
\small
\begin{tikzpicture}
\draw (0,0) -- (8.0,0);
\draw (0,-0.1) -- (0,0.1);
\draw (1,-0.1) -- (1,0.1);
\draw (3.5,-0.1) -- (3.5,0.1);
\draw (4.5,-0.1) -- (4.5,0.1);
\draw (5.5,-0.1) -- (5.5,0.1);
\draw (7.0,-0.1) -- (7.0,0.1);
\draw (8.0,-0.1) -- (8.0,0.1);
\node[above] at (0.0,-0.55) {$-\delta$};
\node[above] at (1.0,-0.55) {$0$};
\node[above] at (3.5,-0.55) {$x-\delta$};
\node[above] at (4.5,-0.55) {$x$};
\node[above] at (5.5,-0.55) {$x+\delta$};
\node[above] at (7.0,-0.55) {$1$};
\node[above] at (8.0,-0.55) {$1+\delta$};
\draw[arrows=<->, >=stealth]  (0.0,1.05) -- (8.0,1.05);
\draw  (0.0,0.25) -- (0.0,1.15);
\draw  (8.0,0.25) -- (8.0,1.15);
\node[above] at (4.00,1.05) {$\extdomd$};
\draw[arrows=<->, >=stealth]  (0.0,0.45) -- (1.0,0.45);
\draw[arrows=<->, >=stealth]  (1.0,0.45) -- (7.0,0.45);
\draw[arrows=<->, >=stealth]  (7.0,0.45) -- (8.0,0.45);
\draw  (1.0,0.25) -- (1.0,0.55);
\draw  (7.0,0.25) -- (7.0,0.55);
\node[above] at (4.00,0.45) {$\Omega$};
\draw[arrows=<->, >=stealth]  (3.5,-0.75) -- (5.5,-0.75);
\draw  (3.5,-0.65) -- (3.5,-0.85);
\draw  (5.5,-0.65) -- (5.5,-0.85);
\node[below] at (4.5,-0.8) {$H_\delta(x)$};
\node[above] at (0.5,0.45) {$D_0^{\delta}$};
\node[above] at (7.5,0.45) {$D_1^{\delta}$};
\end{tikzpicture}
\caption{Definition of domains for the extended domain approach.}
\label{Fig:Extendeddomains}
\end{figure}

Let $\extdomd = [-\delta,1+\delta] \supset \Omega$ be the domain $\Omega$ augmented by the boundary domains $D_0^\delta = [-\delta,0)$ on the left-hand side and $D_1^\delta = (1,1+\delta]$ on the right-hand side, see Figure~\ref{Fig:Extendeddomains}. \edit{The goal here is to construct an extension of the solution $\uperid$ in $\Omega$ to the whole domain $\extdomd$. In the previous section, the integral in~\eqref{eq:forceboundary} can be viewed as the integral over $[x-\delta,x+\delta]$ obtained by extending $\uperid$ to $\uperid(x)=0$, $\forall x \in D_0^\delta$. The issue with such an extension is that it introduces a term involving the first derivative arising from the discontinuity in $\uperid'$ at $x=0$.} We propose here, given a continuous function $\uperid$ in $\overline{\Omega}$, to consider odd extensions of $\uperid$ with respect to $x=0$ and $x=1$ as follows:
\begin{align}
\label{eq:odd-0}
&\uperid(x) = 2 \uperid(0) - \uperid(-x), &&\forall x \in D_0^\delta, \\
\label{eq:odd-1}
&\uperid(x) = 2 \uperid(1) - \uperid(2-x), &&\forall x \in D_1^\delta.
\end{align}
We first note that \corr{the} above conditions imply that the function $\uperid$ is \corr{continuous at $x=0$ and $x=1$ and satisfies the following properties:
\begin{equation}
\begin{aligned}
& \uperid^{(n)}(x^-) = \uperid^{(n)}(x^+), && \text{for odd $n>0$}, \\
& \uperid^{(n)}(x^-) = -\uperid^{(n)}(x^+), && \text{for even $n>0$}, 
\end{aligned}
\end{equation}
where $x=0$ or $1$, $\uperid^{(n)}(x^-)$ and $\uperid^{(n)}(x^+)$ denote the limits of $\uperid^{n}$ when one approaches $x$ from the left and from the right, respectively.
In other words, the odd derivatives of $\uperid$ are continuous at $x=0$ and $x=1$ while the even derivatives are opposite at those points.}

\edit{
Using the above properties and Taylor expansions of $\uperid$ at $x=0$, one can show, following lengthy calculations, that for $x\in (0,\delta)$:
\begin{equation}
\label{eq:EDM-Integral}
\begin{aligned}
& \int_{x-\delta}^{x+\delta} \kappa \frac{\uperid(y) - \uperid(x)}{|y-x|} dy\\
& = \frac{\kappa \delta^2}{2} \bigg[ 
\bigg( 4\frac{x}{\delta} - \Big( 3 - 2 \ln \frac{x}{\delta} \Big) \frac{x^2}{\delta^2}  \bigg) u''(0^+) 
+ x u'''(0) + \ldots \bigg] \\
& = \frac{\kappa \delta^2}{2} \bigg[ 
\bigg( 4\frac{x}{\delta} - \Big( 3 - 2 \ln \frac{x}{\delta} \Big) \frac{x^2}{\delta^2}  \bigg) u''(x) 
+ \mathcal O(\delta) \bigg]
\end{aligned}
\end{equation}
where we have used the facts that $u''(0^+) = u''(x) - x u'''(x) + \ldots$ and $0< x < \delta$. This is a remarkable result which shows that the leading term is now a term that involves the second derivative. The term containing the first derivative vanishes here thanks to the fact that the first derivative of the extended function is continuous at $x=0$. If one had chosen an even extension, this term would not have cancelled.  Moreover, the peridynamic model converges to the classical linear elasticity model as $\delta$ tends to zero if the material property $\kappa$ is replaced, for $x\in (0,\delta)$, by:
\begin{equation}
\label{eq:EDM-correction-left}
\bar{\kappa}(x) = \kappa \bigg( 4\frac{x}{\delta} - \Big( 3 - 2 \ln \frac{x}{\delta} \Big) \frac{x^2}{\delta^2}  \bigg)^{\! -1},
\end{equation}
and, similarly for $x \in (1-\delta,1)$, by:
\begin{equation}
\label{eq:EDM-correction-right}
\bar{\kappa}(x) = \kappa \bigg( 4\frac{1-x}{\delta} - \Big( 3 - 2 \ln \frac{1-x}{\delta} \Big) \frac{(1-x)^2}{\delta^2}  \bigg)^{\! -1}.
\end{equation}
However, the convergence is only linear in $\delta$. As an example, if $x/\delta = 1/2$, one has $\bar{\kappa} \approx k/0.9$, that is $\bar{\kappa} \approx 1.1 \kappa$.}

We now formulate the \corr{peridynamic} problem for the extended domain method. \edit{For the sake of simplicity in the presentation, we shall keep the value of $\kappa$ constant over the whole domain, just keeping in mind that its value can be corrected in the intervals $(0,\delta)$ and $(1-\delta,1)$ according to~\eqref{eq:EDM-correction-left} and~\eqref{eq:EDM-correction-right}, respectively}. The problem consists then in finding \corr{$\uperid \in \extdomd$} such that:
\begin{align}
\label{eq:peridynamicsproblem1}
- \int_{x-\delta}^{x+\delta} \kappa \frac{\uperid(y) - \uperid(x)}{|y-x|} dy & = f_b(x), \quad \forall x \in \Omega,\\
\label{eq:odd0}
\uperid(x) - 2 \uperid(0) + \uperid(-x) & = 0,\qquad \forall x \in D_0^\delta, \\ 
\label{eq:odd1}
\uperid(x) - 2 \uperid(1) + \uperid(2-x) & = 0,\qquad \forall x \in D_1^\delta, \\
\uperid & = 0,\qquad \text{at}\ x=0, \\
\label{eq:NeumannPeri}
\frac{\kappa \delta^2}{2} \uperid' &= g,\qquad \text{at}\ x=1,
\end{align}
where~\eqref{eq:kappa} has been used to modify the Neumann boundary condition~\eqref{eq:Neumann} to~\eqref{eq:NeumannPeri}. The fact that the displacement is extended in the boundary domains by odd extensions allows one to prescribe Dirichlet or Neumann boundary conditions. Indeed, if 
the Neumann condition at $x=1$ is replaced by a non-homogeneous Dirichlet boundary condition, one would simply substitute~\eqref{eq:NeumannPeri} by:
\[
\uperid = \uperid_1,\ \text{at}\ x=1,
\]
where $\uperid_1 \in \mathbb R$ is a prescribed displacement.

Although the solution to the problem~\eqref{eq:peridynamicsproblem1}-\eqref{eq:NeumannPeri} converges to the solution of the local linear elasticity model as the horizon $\delta$ approaches zero, the formulation nevertheless presents several drawbacks:
\begin{enumerate}[label=\Alph*]
\item 
the problem is solved on a larger domain than the domain occupied by the bar, which increases the number of degrees of freedom for a given discretization parameter $h$. This is in particular true when $\delta/h$ is large and in higher dimensions;
\item 
odd extension of the displacement field outside of $\Omega$ may become difficult to carry on in two and three dimensions for domains with complex geometries, especially when boundaries involve corners and edges;
\item
conceptually, there may still be an ambiguity in the application of the Neumann boundary condition, which is derived from the local model, while the behavior in the vicinity of the boundary is governed by the non-local model.
\end{enumerate}
The approach proposed in Section~\ref{Sect:variable-horizon} is an attempt at circumventing those issues.

\subsection{Discrete formulation}

We describe here the discretization of Problem~\eqref{eq:peridynamicsproblem1}-\eqref{eq:NeumannPeri} using a collocation approach. For a given $\delta$, we introduce a uniform grid spacing $h$ chosen such that, as it is customarily done in the literature~\cite{Silling-Askari-CS-2005,Parks-Lehoucq-CPC-2008}, $\delta$ is a multiple of $h$, i.e.\ $\delta/h = m$, with $m$ a positive integer. For the sake of simplicity, but without loss of generality, we shall choose here $m=2$. In this case, choosing $h=1/n$, with $n$ \corr{a} positive integer, the discretization of domain $\extdomd$ is determined by the grid points:
\begin{equation}
x_i = i h, \quad i = -2,-1,0, 1 \ldots, n, n+1, n+2,
\end{equation}
where the points $x_{-2}$, $x_{-1}$, $x_0$ correspond to the grid in \corr{$\overline{D_0^\delta}$} and the points $x_{n}$, $x_{n+1}$, $x_{n+2}$ to the one in \corr{$\overline{D_1^\delta}$}. Moreover, let $u_i$, $i=-2,-1,\ldots,n+2$, denote the discrete displacement at the grid points $x_i$.

Approximation of the integral in~\eqref{eq:peridynamicsproblem1} is obtained by classical quadrature formula using the grid points $x_i$. Note that other more advanced quadrature rules, especially developed for non-local models, could alternatively be used, see e.g.~\cite{Seleson-CMAME-2014,Trask-You-CMAME-2019}. \corr{We then have, using the second-order trapezoidal integration rule,}
\begin{equation}
\label{eq:peridynamics2ndorder}
\begin{aligned}
&\frac{\kappa \delta^2}{2h^2} \left[ - \frac{1}{8} u_{i-2} - \frac{1}{2} u_{i-1} + \frac{5}{4} u_i - \frac{1}{2} u_{i+1} - \frac{1}{8} u_{i+2} \right] \\
&\hspace{1.25in}= f_b(x_i),\quad i = 1,\ldots,n-1,
\end{aligned}
\end{equation}
or, in a more compact form, with $\alpha = {\kappa \delta^2}/(16h^2)$,
\begin{equation}
\label{eq:equationdelta2h}
\begin{aligned}
& - \alpha u_{i-2} - 4 \alpha u_{i-1} + 10 \alpha u_i - 4 \alpha u_{i+1} - \alpha u_{i+2} \\
&\hspace{1.25in}= f_b(x_i),\quad i = 1,\ldots,n-1.
\end{aligned}
\end{equation}
The constraints~\eqref{eq:odd0} and~\eqref{eq:odd1} to enforce odd extensions on the displacements read:
\begin{align}
\label{eq:oddminus2}
&u_{-2} - 2 u_0 + u_{2} = 0, \\
\label{eq:oddminus1}
&u_{-1} - 2 u_0 + u_{1} = 0, \\
\label{eq:oddplus1}
&u_{n-1} - 2 u_{n} + u_{n+1} = 0, \\
\label{eq:oddplus2}
&u_{n-2} - 2 u_{n} + u_{n+2} = 0,
\end{align}
while the Dirichlet boundary condition at $x=0$ implies:
\[
u_0 = 0.
\] 
There remains to discretize the Neumann boundary condition at $x=1$. Several options based on the finite difference method are available. However, since the function $\uperid$ is only $C^1$ at $x=1$, we cannot consider a second-order centered difference of the first derivative. 
We therefore opt for the second-order one-sided finite difference:
\[
\uperid'(1) = \frac{u_{n-2} - 4u_{n-1} +3 u_{n}}{2h} + \mathcal O(h^2),
\]
that, in order to preserve the symmetric feature of the peridynamic model, can be rewritten using~\eqref{eq:oddplus1} and~\eqref{eq:oddplus2} in the centered form as:
\[
\begin{aligned}
\uperid'(1) 
&\approx \frac{2u_{n-2} - 8u_{n-1} + 6 u_{n}}{4h}\\
&\approx \frac{u_{n-2} - 4 u_{n-1} + 6 u_{n} - 4 u_{n-1} + u_{n-2}}{4h}\\
&\approx \frac{u_{n-2} - 4 u_{n-1} + 4 u_{n+1} - u_{n+2}}{4h}.
\end{aligned}
\]
We emphasize that this is a second-order finite difference of $u'(1)$, whether $u$ satisfies the odd extension or not.
We propose to employ the latter formula to discretize the Neumann boundary condition, i.e.\
\[
\frac{\kappa \delta^2}{2} \left[ \frac{ u_{n-2} - 4 u_{n-1} + 4 u_{n+1} - u_{n+2}}{4h} \right] = g,
\]
which can be rewritten as:
\[
2 \alpha h u_{n-2} - 8\alpha h u_{n-1} + 8\alpha h u_{n+1} - 2 \alpha h u_{n+2} = g
\]
with $\alpha={\kappa \delta^2}/(16h^2)$ as before.

The resulting system of linear equations associated with the discretization of the extended domain method is summarized in Figure~\ref{Fig:systemextapproach}. 
\corr{Upon inspection, it is worth noting that the use of the Dirichlet boundary condition at $x=0$ allows one to eliminate the first equation of the system as the value of the degree of freedom $u_{-2}$, associated with the grid point $x_{-2}$ in the extended domain, does not influence the other degrees of freedom. The same conclusion would also hold for $u_{n+2}$ if one had considered a Dirichlet boundary condition at $x=1$. We actually present in the next subsection an approach to further reduce the system of equations.}

\begin{figure*}[ht]
\[
\left[
\begin{array}{rrrrrrrrrrr}
1 & 0 & -2 & 0 & 1 & 0 & 0 & 0 \\ 
0 & 1 & -2 & 1 & 0 & 0 & 0 & 0 \\
0 & 0 & 1 & 0 & 0 & 0 & 0 & 0 & \cdots \\
0 & -\alpha & -4\alpha & 10 \alpha & -4\alpha & -\alpha & 0 & 0 \\
0 & 0 & -\alpha & -4\alpha & 10 \alpha & -4\alpha & -\alpha & 0 \\
& & \vdots & & & \ddots \\
& & & 0 & -\alpha & -4\alpha & 10 \alpha & -4\alpha & -\alpha & 0 & 0 \\
& & & 0 & 0 &-\alpha & -4\alpha & 10 \alpha & -4\alpha & -\alpha & 0 \\
& & & 0 & 0 & 0 & 2\alpha h & -8\alpha h & 0 & 8\alpha h & -2 \alpha h \\
& & & 0 & 0 & 0 & 0 & 1 & -2 & 1 & 0 \\
& & & 0 & 0 & 0 & 1 & 0 & -2 & 0 & 1 \\
\end{array}
\right]
\left[
\begin{array}{l}
u_{-2} \\ u_{-1} \\ u_0 \\ u_1 \\ u_2 \\ \vdots \\ u_{n-2} \\ u_{n-1} \\ u_n \\ u_{n+1} \\ u_{n+2}
\end{array}
\right]
=
\left[
\begin{array}{l}
0 \\ 0 \\ 0 \\ f_b(x_1) \\ f_b(x_2) \\ \vdots \\ f_b(x_{n-2}) \\ f_b(x_{n-1}) \\ g \\ 0 \\ 0
\end{array}
\right]
\]
\caption{System of linear equations associated with the discretization of the extended domain method where $\displaystyle \alpha=\frac{\kappa \delta^2}{16h^2}$.}
\label{Fig:systemextapproach}
\end{figure*}

\subsection{Reduced system of equations}

In fact, it is possible in the one-dimensional case to implicitly enforce the odd extension constraints so as to reduce the system of equations and avoid solving for $u_{-2}$, $u_{-1}$, $u_{n+1}$, and $u_{n+2}$.  Using~\eqref{eq:oddminus1} and~\eqref{eq:oddplus1}, the fourth and $(n-1)$-th equations in the system of Figure~\ref{Fig:systemextapproach} become:
\begin{align}
\label{eq:rightsidedapproximation}
&\frac{\kappa \delta^2}{2} \bigg[ \frac{ - 6 u_0 + 11 u_1 - 4 u_2 - u_3 }{8h^2} \bigg]= f_b(x_1), \\
\label{eq:leftsidedapproximation}
&\frac{\kappa \delta^2}{2} \bigg[ \frac{ - u_{n-3} -4 u_{n-2} + 11 u_{n-1} - 6u_n }{8h^2} \bigg]= f_b(x_{n-1}),
\end{align}
whereas using the left-sided second-order approximation of $u'(1)$ to discretize the Neumann boundary condition gives:
\[
\frac{\kappa \delta^2}{2} \bigg[ \frac{ u_{n-2} - 4 u_{n-1} + 3 u_{n} }{2h} \bigg] = g.
\]
The reduced system of equations is shown in Figure~\ref{Fig:reducedsystemextapproach}. We emphasize here that the extended domain method and its reduced version provide exactly the same solutions for $i=0,\ldots,n$. We will therefore use the reduced system displayed in Figure~\ref{Fig:reducedsystemextapproach} for solving the problems in the numerical examples. We observe that the matrix has in fact a similar structure as the one obtained using the second-order finite difference method for the discretization of the local model (see system of equations in Figure~\ref{Fig:systemlocalmodel}). We verify below that the equations associated with rows $i=1,\ldots,n-1$ provide second-order approximations in $h$, \edit{and a fortiori in $\delta$ since $h=\delta/2$ here,} of $u''$ at the points $x_i$.

\begin{figure*}[ht]
\[
\left[
\begin{array}{rrrrrrr}
1 & 0 & 0 & 0 & 0 & 0 & \cdots \\
-6\alpha & 11 \alpha & -4\alpha & -\alpha & 0 & 0 \\
-\alpha & -4\alpha & 10 \alpha & -4\alpha & -\alpha & 0 \\
\vdots & & & \ddots \\
& 0 & -\alpha & -4\alpha & 10 \alpha & -4\alpha & -\alpha \\
& 0 & 0 &-\alpha & -4\alpha & 11 \alpha & -6\alpha \\
& 0 & 0 & 0 & 4\alpha h & -16\alpha h & 12\alpha h \\
\end{array}
\right]
\left[
\begin{array}{l}
u_0 \\ u_1 \\ u_2 \\ \vdots \\ u_{n-2} \\ u_{n-1} \\ u_n 
\end{array}
\right]
=
\left[
\begin{array}{l}
0 \\ f_b(x_1) \\ f_b(x_2) \\ \vdots \\ f_b(x_{n-2}) \\ f_b(x_{n-1}) \\ g 
\end{array}
\right]
\]
\caption{Reduced system of linear equations associated with the discretization of the extended domain method where $\displaystyle \alpha=\frac{\kappa \delta^2}{16h^2}$.}
\label{Fig:reducedsystemextapproach}
\end{figure*}

Starting with~\eqref{eq:peridynamics2ndorder}, we see that the expression in brackets, multiplied by the factor $1/h^2$, can be rewritten as:
\begin{equation}
\label{eq:approx-secondderivative}
\begin{aligned}
& \frac{ - u_{i-2} - 4 u_{i-1} + 10 u_i - 4 u_{i+1} - u_{i+2} }{8h^2} \\
& \hspace{0.2in} = - \frac{1}{2} \bigg[ \frac{ u_{i-2} - 2 u_i + u_{i+2} }{(2h)^2} \bigg] - \frac{1}{2} \bigg[ \frac{ u_{i-1} - 2 u_i +  u_{i+1} }{h^2} \bigg] \\
& \hspace{0.2in} = - \frac{1}{2} \bigg[ u''(x_i) + \mathcal O ((2h)^2) \bigg] - \frac{1}{2} \bigg[ u''(x_i) + \mathcal O (h^2) \bigg] \\
& \hspace{0.2in} = -  u''(x_i) + \mathcal O (h^2).
\end{aligned}
\end{equation}
In other words, the second derivative $u''(x_i)$ is approximated by the weighted average of the second-order centered differences of the second derivative using, on one hand, the nearest neighbors and, on the other hand, the next-to-nearest neighbors.

We now turn our attention to~\eqref{eq:rightsidedapproximation}. The term in brackets, using classical Taylor expansions, becomes:
\begin{equation}
\label{eq:approx-secondderivative-87}
\begin{aligned}
\frac{ - 6 u_0 + 11 u_1 - 4 u_2 - u_3 }{8h^2} 
&= - \frac{7}{8} u''(x_1) + \mathcal O (h),
\end{aligned}
\end{equation}
where the details of the calculation have been omitted. \corr{This result was to be expected in view of~\eqref{eq:EDM-Integral}}. 
\corr{Since $x_1/\delta = 1/2$ here, the factor $7/8=0.875$ actually corresponds to an approximation of the factor $\kappa /\bar{\kappa}(x_1) \approx 0.9$ that we obtained from~\eqref{eq:EDM-correction-left}. The discrepancy is simply an error due to numerical integration. 
One could multiply the expression by the factor $8/7\approx 1.14$ in order to recover $-u''(x_1)$, or consider the material property at that point to be $\bar{\kappa}(x_1) \approx 1.1\kappa$ according to~\eqref{eq:EDM-correction-left}, but we note that the approach would still remain suboptimal due to the fact that one gets only a first-order approximation.} Note that the same conclusion holds in the case of~\eqref{eq:leftsidedapproximation}. We will illustrate the spurious perturbations near the boundaries in the numerical examples depending on the chosen correction method.

\section{Variable horizon method}
\label{Sect:variable-horizon}

\subsection{Continuous formulation}

\begin{figure}
\centering
\small
\begin{tikzpicture}
\draw[arrows=->, >=stealth] (1,0) -- (9.0,0);
\draw[arrows=->, >=stealth] (1,0) -- (1,2.0);
\draw (1.0,0.0) -- (2.5,1.5);
\draw (2.5,1.5) -- (6.5,1.5);
\draw (6.5,1.5) -- (8.0,0.0);
\draw (2.5,0.2) -- (2.5,1.3);
\draw (6.5,0.2) -- (6.5,1.3);
\draw (2.5,-0.1) -- (2.5,0.1);
\draw (6.5,-0.1) -- (6.5,0.1);
\draw (8.0,-0.1) -- (8.0,0.1);
\node[above] at (0.7,1.5) {$\delta_v$};
\node[above] at (1.0,-0.55) {$0$};
\node[above] at (2.5,-0.55) {$\delta$};
\node[above] at (6.5,-0.55) {$1-\delta$};
\node[above] at (8.0,-0.55) {$1$};
\node[above] at (9.0,-0.55) {$x$};
\draw (2.5,0) circle (1.5);
\draw (1.75,0) circle (0.75);
\draw (1.75,-0.1) -- (1.75,0.1);
\draw (1.375,0) circle (0.375);
\draw (1.375,-0.1) -- (1.375,0.1);
\end{tikzpicture}
\caption{Example of a variable horizon function $\delta_v(x)$. The circles centered at points \corr{$x\in B_0^{\delta}=(0,\delta)$} are introduced to represent the associated domain $H_\delta(x)$.}
\label{Fig:variablehorizon}
\end{figure}
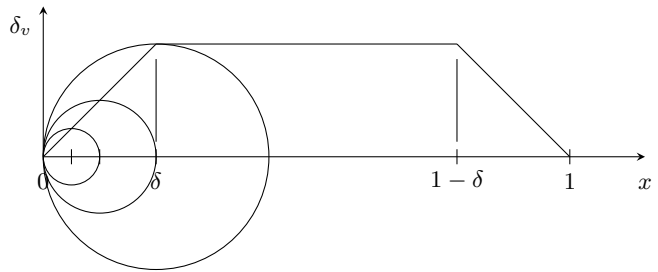

Our motivation in this approach is to address some of the ambiguities associated with the extended domain approach, the first one being that the problem is defined on a larger domain than the original domain~$\Omega$ occupied by the bar (although we have shown than the problem from the extended domain method can be recast, in the one-dimensional case, as a reduced problem defined on~$\Omega$). Our objective is therefore to introduce a new problem restricted to the computational domain~$\Omega$. \corr{Secondly, in order to avoid constructing corrected forces near the boundaries, the set $H_\delta(x)$ should always be a subset of domain $\Omega$, suggesting that the horizon should tend to zero as one approaches the boundary}. Finally, we suppose that one is interested in using a non-local model with constant horizon $\delta$ in the interior of $\Omega$ sufficiently far from the boundaries. In order to satisfy all these requirements, one needs to construct a non-local model with variable horizon $\delta_v(x)$ such that, $\forall x \in \Omega$:
\begin{equation}
\begin{aligned}
& 0 \leq \delta_v(x) \leq \delta, \\
& x - \delta_v(x) \geq 0, \\
& x + \delta_v(x) \leq 1.
\end{aligned}
\end{equation} 
The definition of the function $\delta_v(x)$ is therefore not unique. However, the simplest continuous function that fulfill \corr{the} above requirements is the piecewise linear function defined as:
\begin{equation}
\label{eq:deltafn}
\delta_v(x) = \left\{ 
\begin{array}{ll} 
x, & \quad 0 < x \leq \delta, \\ 
\delta, & \quad \delta < x \leq 1 - \delta, \\ 
1 - x, & \quad 1-\delta < x < 1. 
\end{array}
\right.
\end{equation}
The function $\delta_v(x)$ is shown in Figure~\ref{Fig:variablehorizon}. Alternative functions, in particular smoother functions, could also be considered as long as they \corr{take values} between $0$ and $\delta_v(x)$ (as defined in~\eqref{eq:deltafn}) for all $x \in \Omega$. In view of~\eqref{eq:kappa}, the horizon being a function of $x$ implies that the material parameter $\kappa$ should also depends on $x$, i.e.\ $\kappa=\bar\kappa(x)$, if the non-local model is to be compatible with the local model.
The new problem consists in searching for $\uperid(x)$ such that:
\begin{align}
\label{eq:peridynamicsproblem2}
- \int_{x-\delta_v(x)}^{x+\delta_v(x)} \bar\kappa(x) \frac{\uperid(y) - \uperid(x)}{|y-x|} dy
&= f_b(x), \quad \forall x \in \Omega,\\
\uperid &= 0, \qquad \text{at}\ x=0, \\
\label{eq:NeumannPeri2}
EA \uperid' &= g, \qquad \text{at}\ x=1.
\end{align}
In order for the model to be compatible with the linear elasticity model everywhere in the domain with a convergence of order $\mathcal O(\delta^2)$, the product $\bar\kappa(x) \delta_v^2(x)$ \corr{needs to remain} constant for all $x \in \Omega$, so that 
\begin{equation}
\label{eq:productkappadelta}
\bar\kappa(x) \delta_v^2(x) = \kappa \delta^2, \quad \forall x \in \Omega,
\end{equation}
where $\kappa$ satisfies~\eqref{eq:kappa}. \edit{We note that this scaling of the material parameter appears to be different from that in~\cite{bobaru2009-1D}.}

\subsection{Discrete formulation}
\label{Sect:DFvariablehorizon}

As before, we shall use a collocation approach to discretize the problem~\eqref{eq:peridynamicsproblem2}-\eqref{eq:NeumannPeri2} and choose the parameter $h$ such that $\delta/h = m$, with $m=2$. The grid \corr{in} $\Omega$ is defined by the grid points:
\begin{equation}
x_i = i h,\quad i = 0,1, \ldots, n,
\end{equation}
and $u_i$, $i=0,1,\ldots,n$, denote the discrete displacements at the grid points $x_i$. From~\eqref{eq:deltafn}, we observe that: 
\[
\delta_v(x_i) = 
\left\{ 
\begin{array}{ll} \delta/2 = h, & \quad i = 1, \\ \delta = 2h, & \quad i=2, \ldots,n-2, \\ \delta/2 = h, & \quad  i=n-1. \end{array} \right.
\]

The approximation of~\eqref{eq:peridynamicsproblem2} at $x_{1}$ and $x_{n-1}$ by a second-order integration scheme for the integral is given as:
\[
\frac{\bar\kappa(x_i)}{2} \left[-u_{i-1} + 2u_i - u_{i+1} \right] = f_b(x_i), \quad i=1,n-1.
\]
However, using relation~\eqref{eq:productkappadelta} and the fact that, in those two cases, $\delta_v(x_i) = h$, one can rewrite \corr{the} above equation as:
\[
\frac{\kappa \delta^2}{2h^2} \left[-u_{i-1} + 2u_i - u_{i+1} \right] = f_b(x_i), \quad i=1,n-1,
\]
or, using again the coefficient $\alpha =\kappa \delta^2/(16h^2)$, 
\[
- 8\alpha u_{i-1} + 16\alpha u_i - 8 \alpha u_{i+1} = f_b(x_i), \quad i=1,n-1.
\]
In the case of the remaining interior grid points $x_i$, $i=2,\ldots,n-2$, the approximation of~\eqref{eq:peridynamicsproblem2} is given by~\eqref{eq:equationdelta2h}, that is:
\[
\begin{aligned}
& - \alpha u_{i-2} - 4 \alpha u_{i-1} + 10 \alpha u_i - 4 \alpha u_{i+1} - \alpha u_{i+2} \\
&\hspace{1.25in}= f_b(x_i),\quad i = 2,\ldots,n-2.
\end{aligned}
\]
The Dirichlet boundary condition is prescribed as usual as:
\[
u_0 = 0.
\]
Finally, it remains to discretize the Neumann boundary condition. A second-order approximation of the first derivative using a left-sided finite difference scheme is:
\[
\uperid' (1) = \frac{u_{n-2} - 4 u_{n-1} + 3u_n}{2h} + \mathcal O(h^2).
\] 
Therefore, the approximation of the Neumann condition, using $EA=\kappa\delta^2/2$ and parameter $\alpha$, reads:
\[
4\alpha h u_{n-2} - 16 \alpha h u_{n-1} + 12 \alpha h u_{n} = g.
\] 

The resulting system of linear equations, corresponding to the discretization of the variable horizon method, is summarized in Figure~\ref{Fig:systemvariablehorizon}. We note here that the coefficient $\alpha$ could be factorized in front of the matrix. \corr{We conclude the section by emphasizing that the variable horizon method provides an approximation of the classical linear elasticity model of order two in the parameter $h$, and a fortiori, in $\delta$. We will see that this rate of convergence is consistently observed in the numerical examples presented below.}

\begin{figure*}
\[
\left[
\begin{array}{rrrrrrrrrrr}
1 & 0 & 0 & 0 & 0 & 0 & \cdots \\
-8\alpha & 16 \alpha & -8\alpha & 0 & 0 & 0 \\
-\alpha & -4\alpha & 10 \alpha & -4\alpha & -\alpha & 0 \\
\vdots & & & \ddots \\
& 0 & -\alpha & -4\alpha & 10 \alpha & -4\alpha & -\alpha \\
& 0 & 0 & 0 & -8\alpha & 16 \alpha & -8\alpha \\
& 0 & 0 & 0 & 4\alpha h & -16\alpha h & 12\alpha h \\
\end{array}
\right]
\left[
\begin{array}{l}
u_0 \\ u_1 \\ u_2 \\ \vdots \\ u_{n-2} \\ u_{n-1} \\ u_n 
\end{array}
\right]
=
\left[
\begin{array}{l}
 0 \\ f_b(x_1) \\ f_b(x_2) \\ \vdots \\ f_b(x_{n-2}) \\ f_b(x_{n-1}) \\ g
\end{array}
\right]
\]
\caption{System of linear equations associated with the discretization of the variable horizon method where $\displaystyle \alpha =\frac{\kappa \delta^2}{16h^2}$.}
\label{Fig:systemvariablehorizon}
\end{figure*}

\section{Numerical examples}
\label{Sect:numericalexamples}

The objective of this section is to present several numerical examples in order to compare the solutions \corr{of} the extended domain and variable horizon approaches for implementing boundary conditions in non-local \corr{peridynamic} models. For completeness, we will also show the results obtained with the local linear elasticity model. The system of equations for the discretization of the problem~\eqref{eq:1dlinearelasticity}-\eqref{eq:Neumann} by a second-order finite difference method is shown in Figure~\ref{Fig:systemlocalmodel}, where we have assumed the same grid on~$\Omega$ than the one defined in Section~\ref{Sect:DFvariablehorizon}. For simplicity, but without loss of generality, we shall consider a one-dimensional bar for which $EA=1$, so that \corr{$\kappa = 2/\delta^2$}, $\alpha = 1/(8h^2)$, and $\beta = 1/(2h^2)$. Moreover, the value of the boundary data for the Neumann boundary condition is set in all examples to unity, i.e.\ $g=1$, unless explicitly said otherwise. \corr{The first series of examples will be concerned with $\delta$-convergence (see e.g.~\cite{bobaru2009-1D,seleson2016convergence}), for which we shall measure the error in the peridynamic solutions with respect to the classical local solution. In this case, we note that the discretization parameter $h$ implicitly goes to zero as well since $h=\delta/2$. We then provide an example to study the effect of the correcting factor in the material property for the extended domain method. The following example will treat the case of $m$-convergence, for which $\delta$ is fixed and $h$ goes to zero, in order to show that the approximate solution actually converges to the corresponding peridynamic solution. Finally, we consider an example whose solution is given in terms of an exponential function and exhibits a very steep gradient at one end point. The objective is to show that, as expected, the rates of convergence are unaltered even in the presence of large variations in the solution.}

\begin{figure*}
\[
\left[
\begin{array}{rrrrrrrrrrr}
1 & 0 & 0 & 0 & 0 & 0 & \cdots \\
-2\beta & 4 \beta & -2\beta & 0 & 0 & 0 \\
0 & -2\beta & 4 \beta & -2\beta & 0  & 0 \\
\vdots & & & \ddots \\
& 0 & 0 & -2\beta & 4 \beta & -2\beta & 0 \\
& 0 & 0 & 0 & -2\beta & 4 \beta & -2\beta \\
& 0 & 0 & 0 & \beta h & -4\beta h & 3\beta h 
\end{array}
\right]
\left[
\begin{array}{l}
u_0 \\ u_1 \\ u_2 \\ \vdots \\ u_{n-2} \\ u_{n-1} \\ u_n 
\end{array}
\right]
=
\left[
\begin{array}{l}
 0 \\ f_b(x_1) \\ f_b(x_2) \\ \vdots \\ f_b(x_{n-2}) \\ f_b(x_{n-1}) \\ g
\end{array}
\right]
\]
\caption{System of linear equations associated with the finite difference discretization of the local linear elasticity model with $\displaystyle \beta =\frac{EA}{2h^2} = 4\alpha$.}
\label{Fig:systemlocalmodel}
\end{figure*}

\subsection{$\delta$-convergence numerical study}

\edit{We consider in this section four examples for which the manufactured solutions are polynomial functions of degree up to four. The objectives of this numerical study are to analyze the rates of convergence in $\delta$ as the peridynamic solutions converge to the linear elasticity solution and to highlight the differences between the methods. For comparison, we compute the maximum relative error:
\[
\mathcal E_n = \max_{i=1,\ldots,n} \left| \frac{\uelast(x_i) - u_i}{\uelast(x_i)} \right|,
\]
with $n$ the total number of degrees of freedom of the numerical solution $u_i$, $i=1,\ldots,n$,
for the local linear elasticity model (LLEM), the extended domain method (EDM) without correction, and the variable horizon method (VHM).}

\subsubsection*{Linear solution}
The first example that we shall consider is defined when the body force density $f_b$ identically vanishes on $\Omega$, i.e. $f_b(x) = 0$, $\forall x \in \Omega$, in which case the solution \corr{to  Equations~\eqref{eq:1dlinearelasticity}-\eqref{eq:Neumann}} is the linear function:
\[
\uelast(x) = gx, \quad \forall x \in \bar{\Omega} = [0,1].
\]
It is not difficult to show that the linear discrete function $u_i=igh$, $i=0,\ldots,n$ is \corr{the} solution to the systems of equations given in \corr{Figures~\ref{Fig:reducedsystemextapproach},~\ref{Fig:systemvariablehorizon}}, and~\ref{Fig:systemlocalmodel}, for any value of $h$ with $h=1/n$. In other words, the \corr{linear function is exactly reproduced} by the discrete solutions of the local linear elasticity model (LLEM), the extended domain method (EDM), and the variable horizon method (VHM). 

\subsubsection*{Quadratic solution}

The second example deals with the problem for which the body force density is given by:
\[
f_b(x) = 1.
\]
In that case, the exact solution to the local model equations~\eqref{eq:1dlinearelasticity}-\eqref{eq:Neumann} reads:
\[
\uelast(x) = \frac{x( 4 - x)}{2}.
\]
We now solve the discrete problems with $\delta=1/2$, $1/4$, $1/8$, and $1/16$, (i.e.\ $h=1/4$, $1/8$, $1/16$, and $1/32$, respectively) and report the maximum relative error in Table~\ref{Tab:quadratic}. We observe on the one hand that the relative errors for LLEM and VHM are identically zero independently of the grid size $h$, as expected. Indeed, since the derivatives of order three and higher of the exact solution $\uelast$ all vanish, the discrete methods are able in \corr{this} case to exactly reproduce the solution.

\begin{table}
\center
\begin{tabular}{ccccc}
& & \multicolumn{3}{c}{Maximum relative error $\mathcal E_n$} \\
\toprule
$n$ & $\delta$ & LLEM & EDM & VHM \\
\midrule
4 & 0.5 & 0 & 0.05098 & 0 \\
8 & 0.25 & 0 & 0.02443 & 0 \\
16 & 0.125 & 0 & 0.01202 & 0  \\
32 & 0.0625 & 0 & 0.00596 & 0  \\
\bottomrule
\end{tabular}
\caption{Maximum relative error $\mathcal E_n$ for the quadratic solution obtained with the local linear elasticity model (LLEM), the extended domain method (EDM), and the variable horizon method (VHM).}
\label{Tab:quadratic}
\end{table}

\begin{figure}
\centering 
\includegraphics[width=0.9\columnwidth]{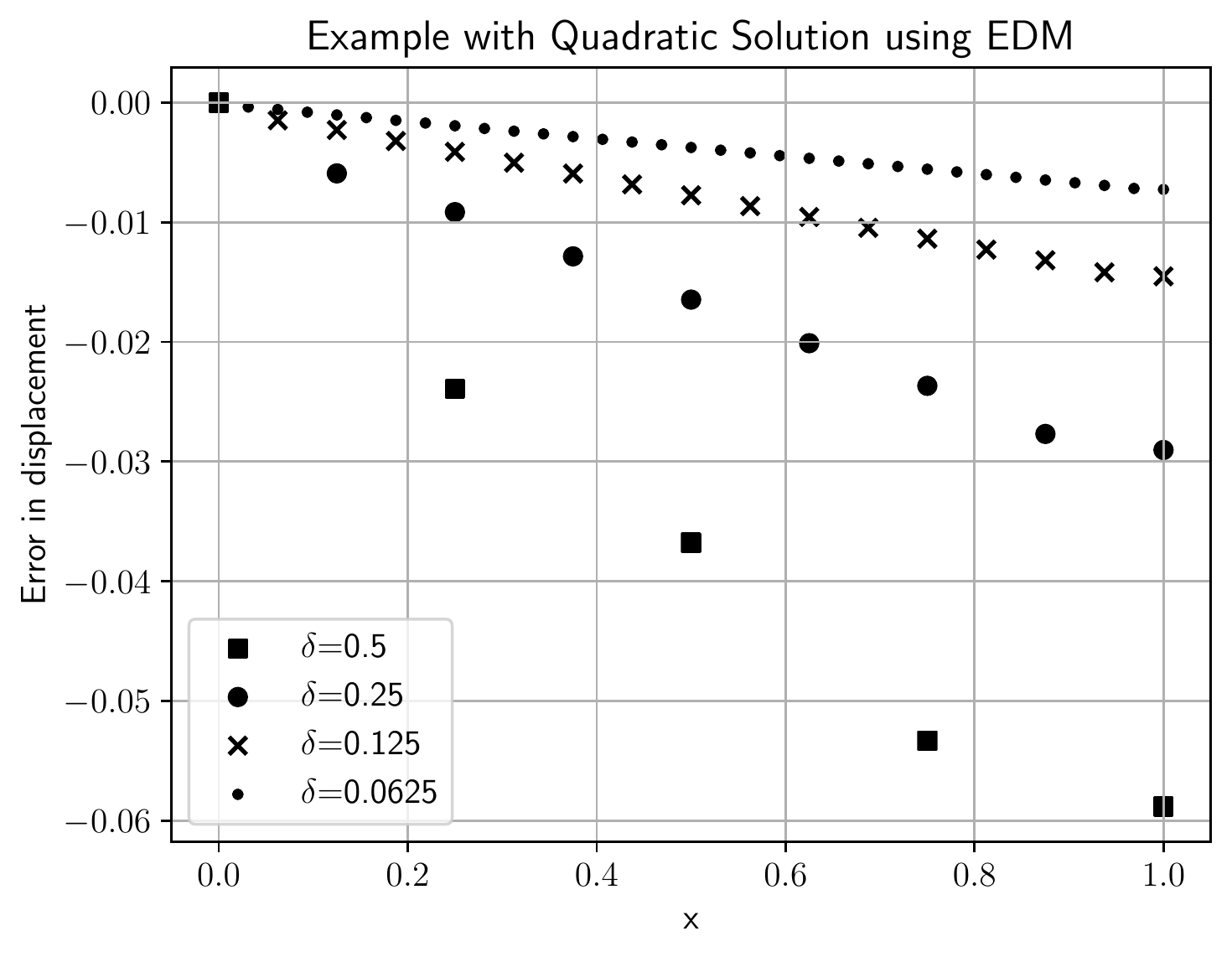}
\caption{Error in displacement for the quadratic solution using the extended domain method (EDM) for various values of horizon~$\delta$.}
\label{Fig:disp-error-quadratic}
\end{figure}

On the other hand, we observe that the errors in the displacement when using EDM are nonzero and seem to decrease with a rate of convergence of order one with respect to~$h$. In this case, the only sources of error arise from the fact that the second derivative is not properly estimated at $x_1 = h$ and $x_{n-1} = 1-h$, as explained by the result in Eq.~\eqref{eq:approx-secondderivative-87}. As a matter of fact, the error becomes zero if we multiply the left-hand side in the second and $(n-1)$-th equations of Figure~\ref{Fig:reducedsystemextapproach} by the correcting factor $8/7$. Finally, we show in Figure~\ref{Fig:disp-error-quadratic} the error $e_i = \uelast(x_i) - u_i$, $i=1,\ldots,n$, in order to illustrate how the sources of error committed at $x_1$ and $x_{n-1}$ \corr{pollute} the solution in the whole computational domain.  

\subsubsection*{Cubic solution}

We repeat the previous experiment with the following body force density:
\[
f_b(x) = x,
\]
for which the exact solution \corr{to Equations~\eqref{eq:1dlinearelasticity}-\eqref{eq:Neumann}} is given by:
\[
\uelast(x) = \frac{x( 3 - x)(3+x)}{6}.
\]
The results are reported in Table~\ref{Tab:cubic}. In the case of LLEM and VHM, we observe that the maximum relative errors decrease with a rate of convergence of order two with respect to $h$, as expected. Moreover, the errors for both methods are identical. This is explained by the fact that the source of error only arises from the approximation of the Neumann boundary condition. Indeed, the approximation being of order two, the leading term of the truncation error involves the third derivative of $\uelast$, which is nonzero, while the leading term of the truncation error for the approximation of the second derivative involves the fourth derivative of $\uelast$, which is in this case zero. Therefore, since the Neumann boundary condition is discretized in the same way for LLEM and VHM, the two methods provide the same approximate solution of the problem, hence the same relative error.  

\begin{table}
\center
\begin{tabular}{ccccc}
& & \multicolumn{3}{c}{Maximum relative error $\mathcal E_n$} \\
\toprule
$n$ & $\delta$ & LLEM & EDM & VHM \\
\midrule
4 & 0.5 & 0.01481 & 0.02774 & 0.01481 \\
8 & 0.25 & 0.00379 & 0.01764 & 0.00379 \\
16 & 0.125 & 0.00096 & 0.00983 & 0.00096 \\
32 & 0.0625 & 0.00024 & 0.00517 & 0.00024 \\
\bottomrule
\end{tabular}
\caption{Maximum relative error $\mathcal E_n$ for the cubic solution obtained with the local linear elasticity model (LLEM), the extended domain method (EDM), and the variable horizon method (VHM).}
\label{Tab:cubic}
\end{table}

\begin{figure}
\centering
\includegraphics[width=0.9\columnwidth]{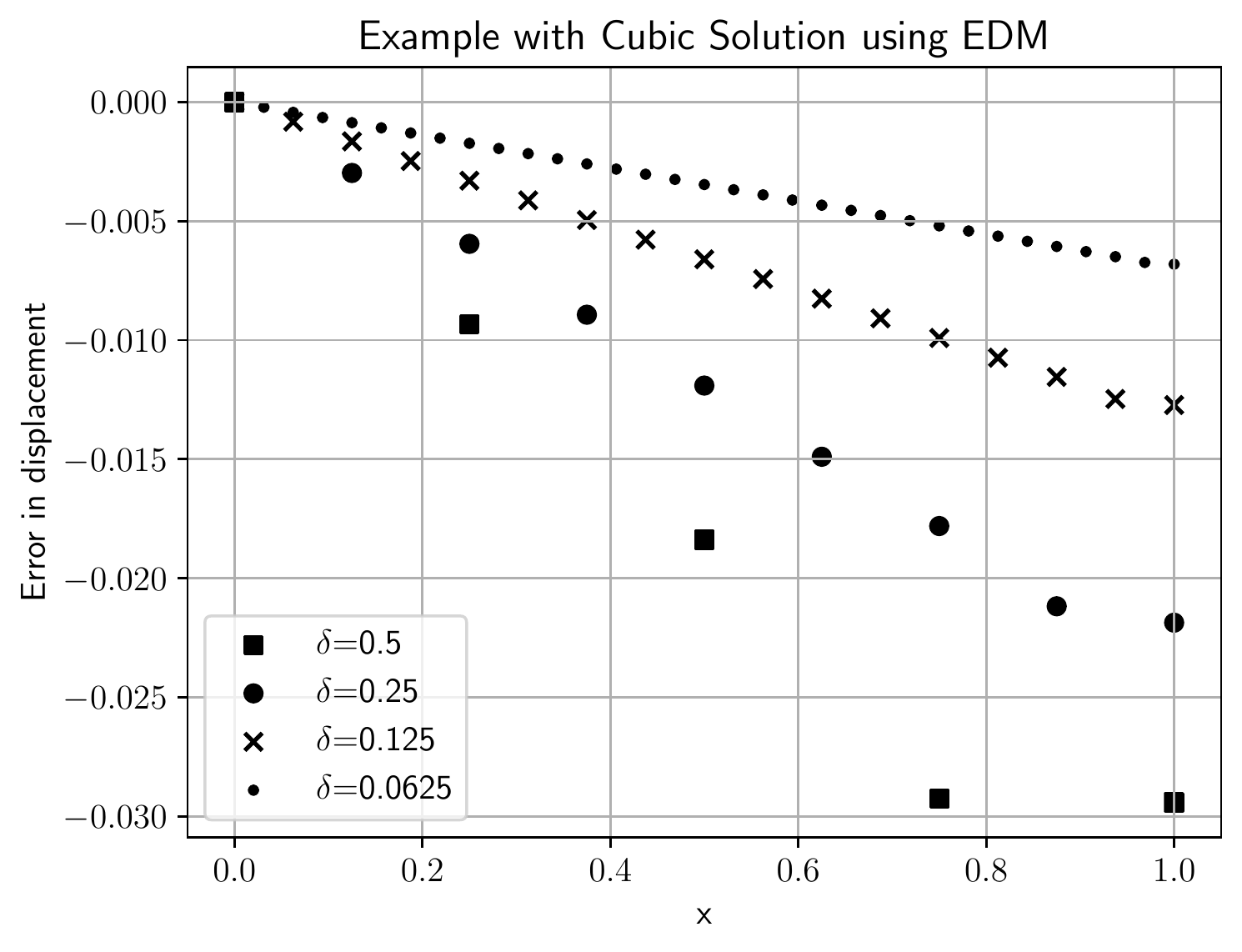}
\caption{Error in displacement for the cubic solution using the extended domain method (EDM) for various values of horizon~$\delta$.}
\label{Fig:disp-error-cubic}
\end{figure}

As far as EDM is concerned, we note the same rate of convergence, approximately of order one, as in the previous problem with the quadratic solution. It is once again obvious from Figure~\ref{Fig:disp-error-cubic} that the issue comes from the approximation of the second derivative at $x_1$ and $x_{n-1}$. \corr{Indeed, we observe a clear change in the slope of the displacement error at $x_{n-1}$. The issue is however less visible at $x_1$ as the solution is close to zero near the boundary.}

\subsubsection*{Quartic solution}

We repeat the previous experiment and consider the case of the quadratic body force density:
\[
f_b(x) = x^2,
\]
which provides the following quartic solution \corr{to Equations~\eqref{eq:1dlinearelasticity}-\eqref{eq:Neumann}}:
\[
\uelast(x) = \frac{x(16-x^3)}{12}.
\]
The maximum relative errors are shown in Table~\ref{Tab:quartic}. We first note that the relative error for both LLEM and VHM decreases with a rate of convergence of order two similarly to the previous example. However, the errors in the solutions obtained with VHM are slightly greater than those in the solutions obtained by LLEM. Since the fourth derivative of $\uelast$ is no longer zero in $\Omega$, the error is now partly due to the approximation of the second derivative. In the case of VHM, it is obtained as the average of two centered difference stencils of order $\mathcal O(h^2)$ and $\mathcal O((2h)^2)$, respectively (see e.g.\ Eq.~\eqref{eq:approx-secondderivative}) at the grid points $x_i$, $i=2,\ldots,n-2$. It naturally follows that the solution is slightly less accurate than the solution obtained from LLEM, which only involves approximations of order $\mathcal O(h^2)$. \corr{However, the difference between the errors in the two solutions correspond to the modeling error of order $\delta^2$ that arises from using the non-local model instead of the local model.} 

\begin{table}
\center
\begin{tabular}{ccccc}
& & \multicolumn{3}{c}{Maximum relative error $\mathcal E_n$} \\
\toprule
$n$ & $\delta$ & LLEM & EDM & VHM \\
\midrule
4 & 0.5 & 0.03226 & 0.00663 & 0.03617 \\
8 & 0.25 & 0.00891 & 0.01267 & 0.01085 \\
16 & 0.125 & 0.00233 & 0.00889 & 0.00294 \\
32 & 0.0625 & 0.00060 & 0.00511 & 0.00076 \\
\bottomrule
\end{tabular}
\caption{Maximum relative error $\mathcal E_n$ for the quartic solution obtained with the local linear elasticity model (LLEM), the extended domain method (EDM), and the variable horizon method (VHM).}
\label{Tab:quartic}
\end{table}

\begin{figure}
\centering
\includegraphics[width=0.9\columnwidth]{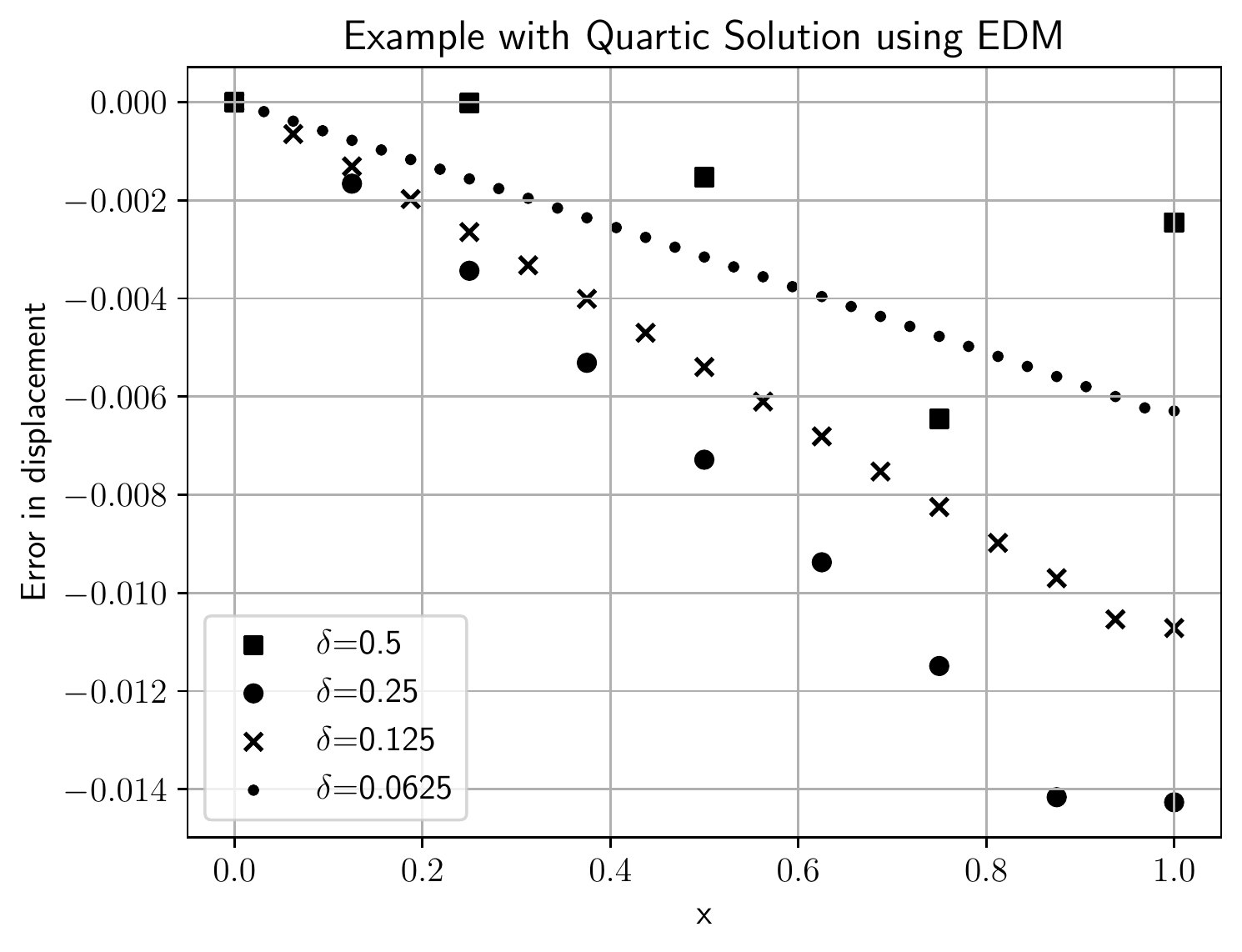}
\caption{Error in displacement for the quartic solution using the extended domain method (EDM) for various values of horizon~$\delta$.}
\label{Fig:disp-error-quartic-EDM}
\end{figure}

\begin{figure}
\centering
\includegraphics[width=0.9\columnwidth]{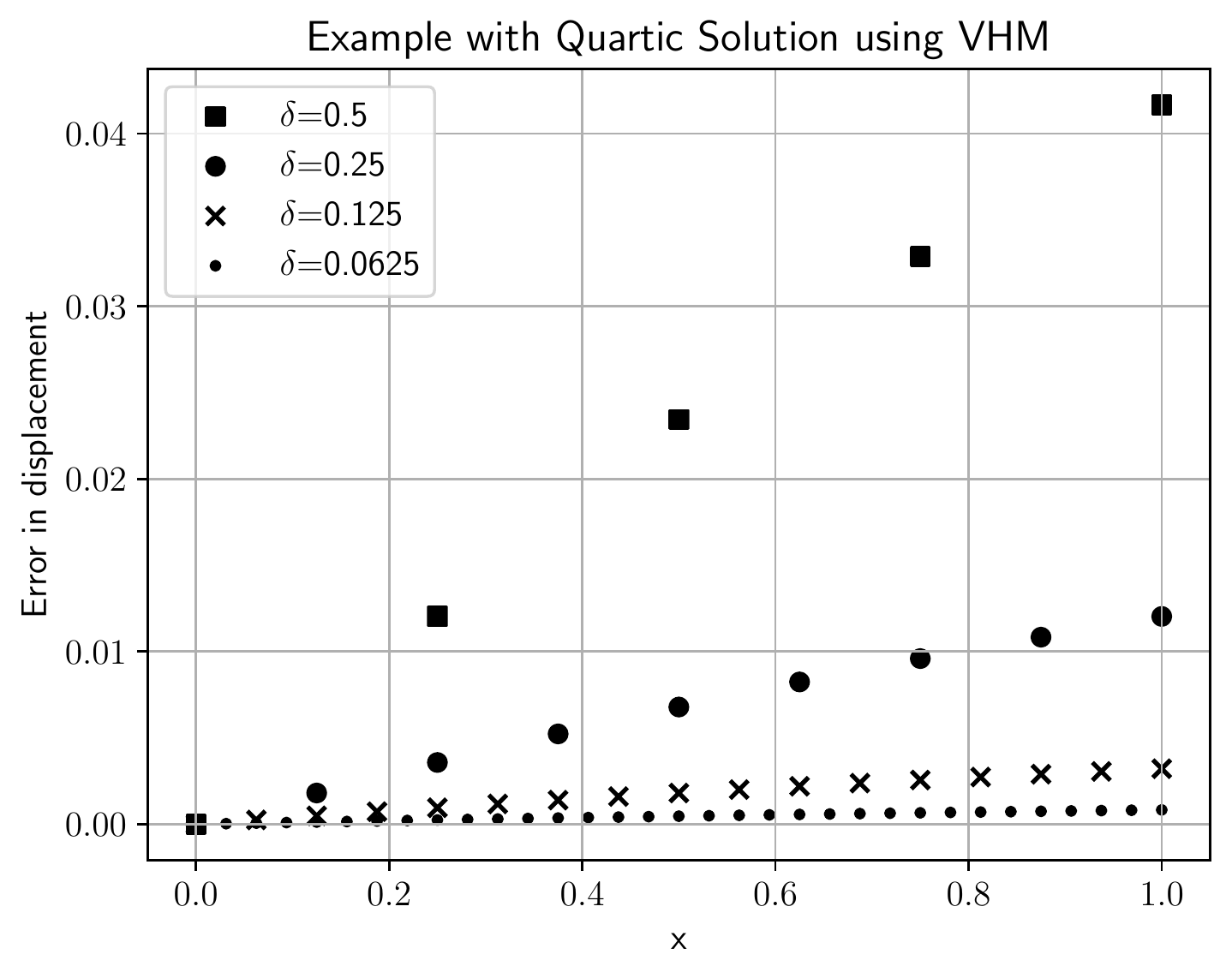}
\caption{Error in displacement for the quartic solution using the variable horizon method (VHM) for various values of horizon~$\delta$.}
\label{Fig:disp-error-quartic-VHM}
\end{figure}

Finally, we show in Figure~\ref{Fig:disp-error-quartic-EDM} the displacement error for the EDM solution. The conclusions that we drew in the cases of the quadratic and cubic solutions also hold, in addition to the fact that the behavior of the error seems even more unpredictable here. For comparison, we show in Figure~\ref{Fig:disp-error-quartic-VHM} the displacement error in the VHM solution, which exhibits a monotonic and smoother behavior and decreases much faster as $\delta$ goes to zero.

\subsection{Effect of correction factor in EDM}

\begin{figure}
    \centering
    \includegraphics[width=0.9\columnwidth]{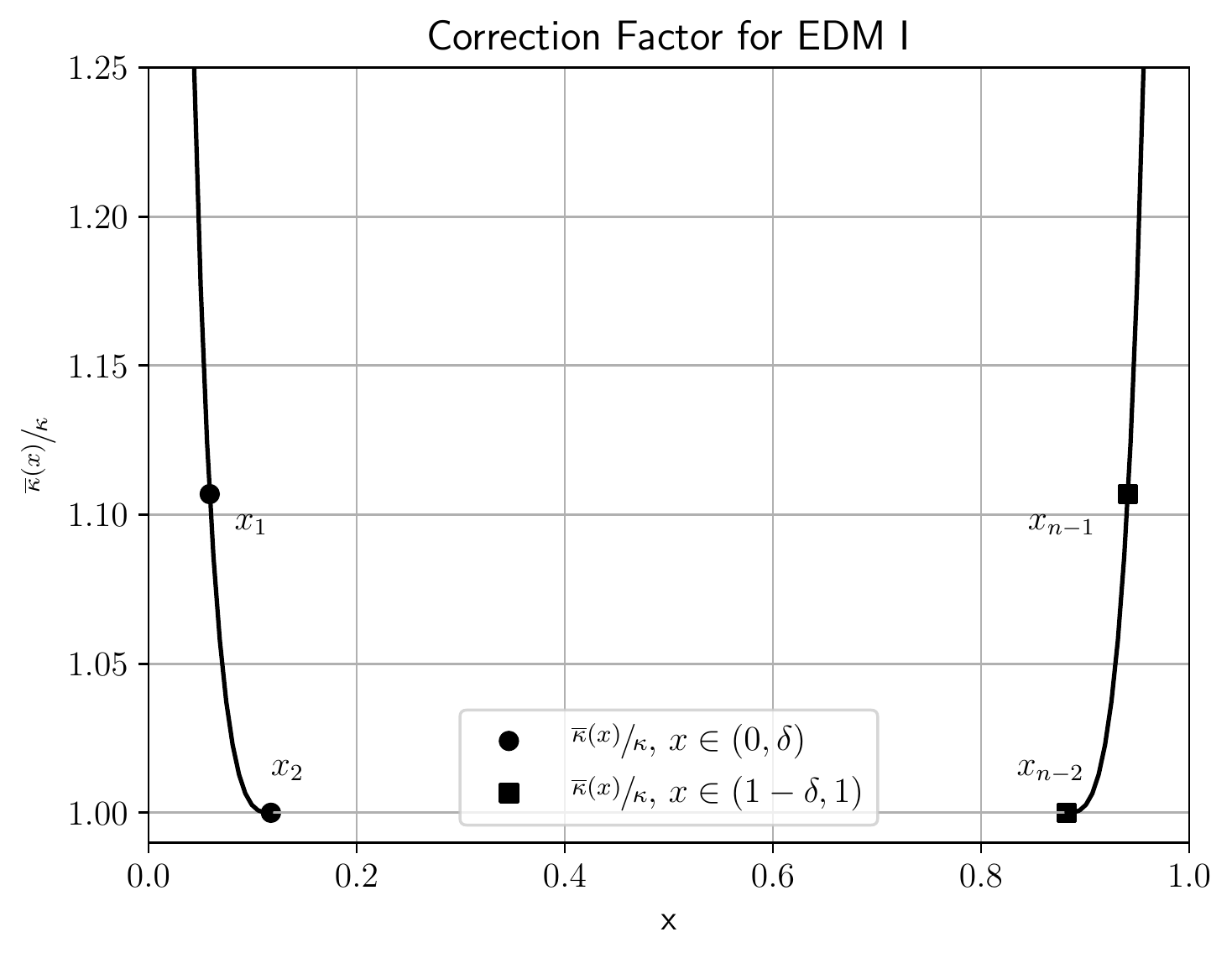}
    \caption{Plot of the correction factor ${\overline{k}(x)}/{\kappa}$ provided by~\eqref{eq:EDM-correction-left} on the left-hand side and by~\eqref{eq:EDM-correction-right} on the right-hand side for $\delta=0.125$.}
    \label{fig:correctionfactorEDM}
\end{figure}

\begin{figure}
    \centering
    \includegraphics[width=0.9\columnwidth]{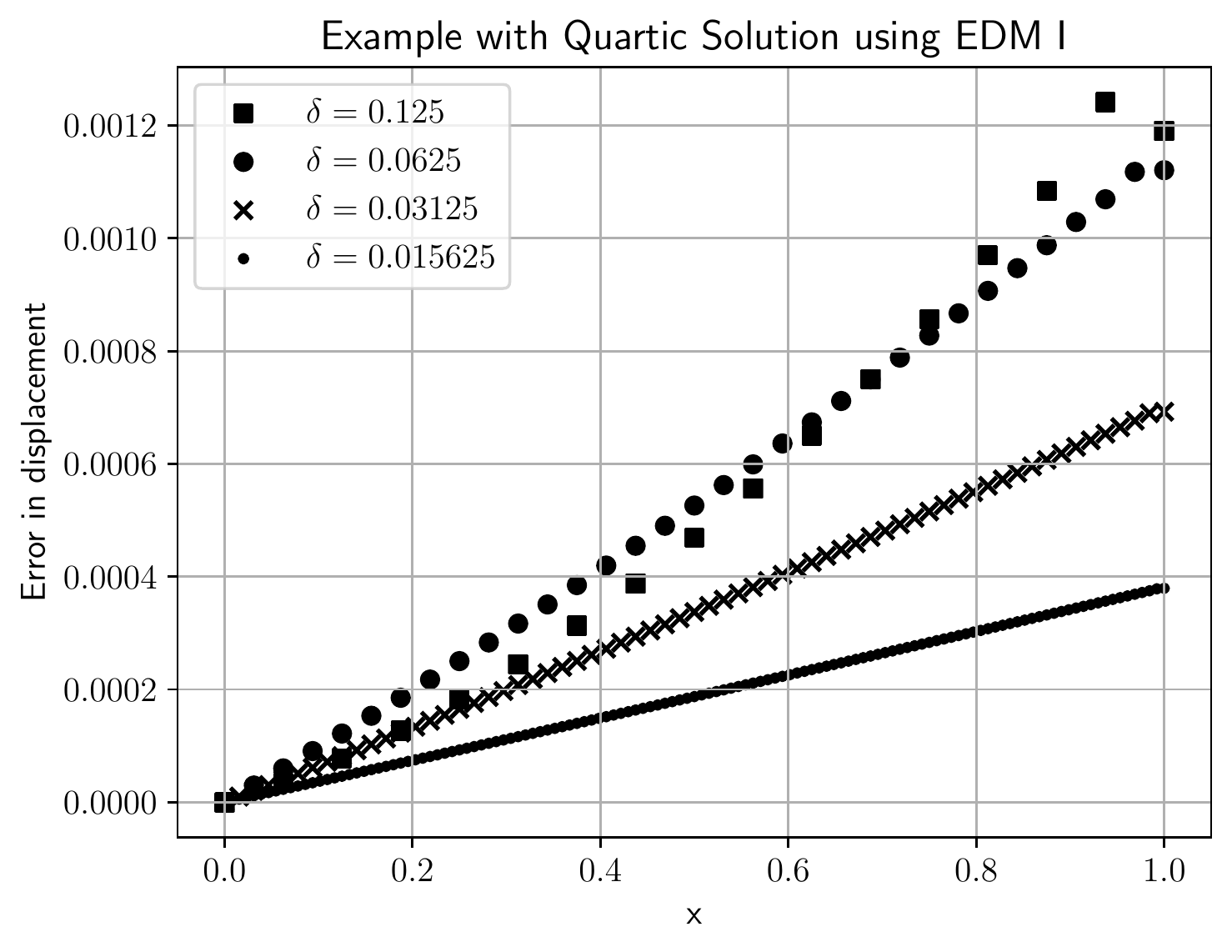}
    \caption{Error in displacement using EDM I with correction factor ${\overline{k}(x)}/{\kappa}$ given by~\eqref{eq:EDM-correction-left} and~\eqref{eq:EDM-correction-right} for various values of the horizon~$\delta$.}
    \label{fig:EDMI}
\end{figure}

\begin{figure}
    \centering
    \includegraphics[width=0.9\columnwidth]{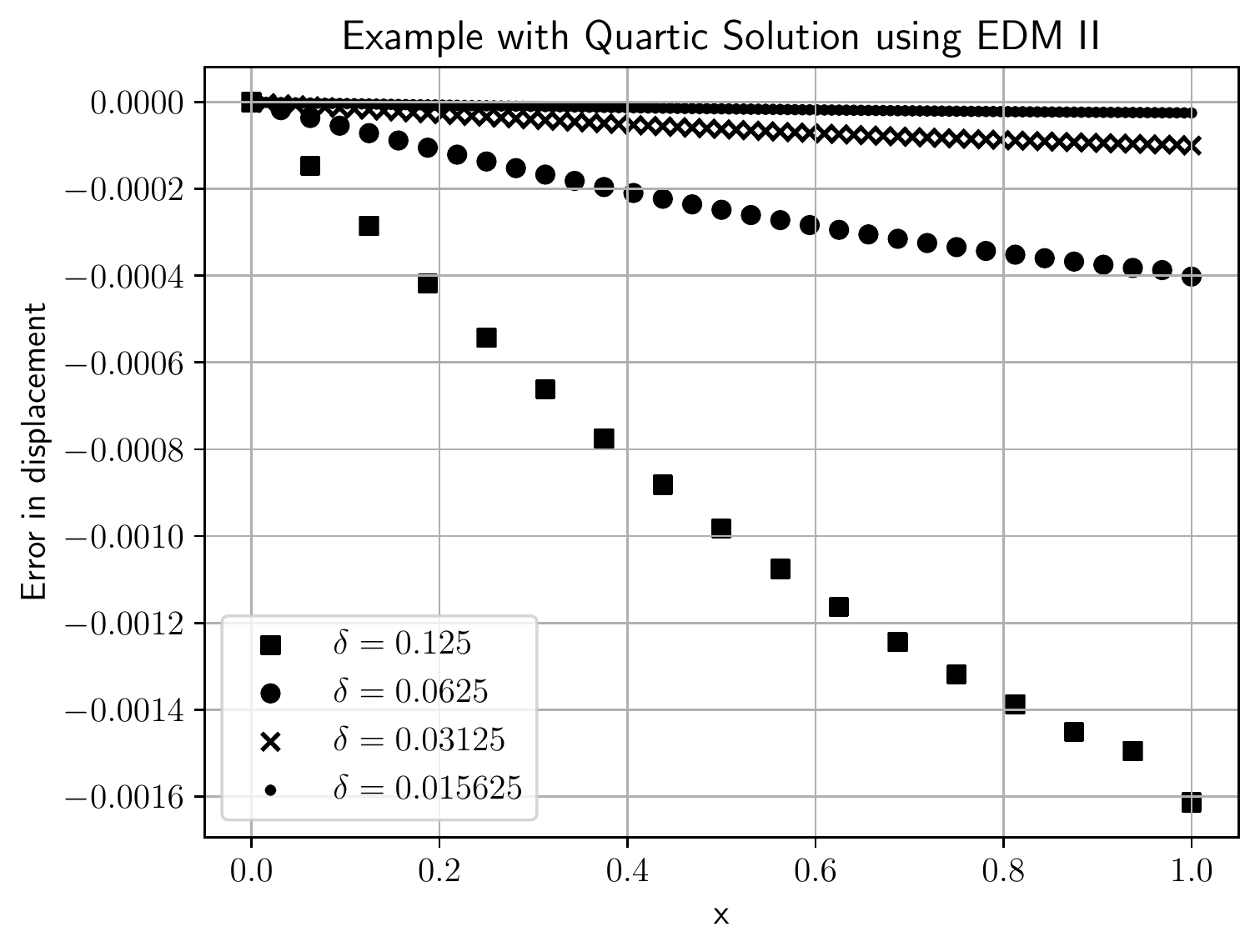}
    \caption{Error in displacement using EDM II with correction factor $\overline{k}(x_1)/\kappa=\overline{k}(x_{n-1})/\kappa=8/7$ for various values of the horizon~$\delta$.}
    \label{fig:EDMII}
\end{figure}

\begin{figure}
    \centering
    \includegraphics[width=0.9\columnwidth]{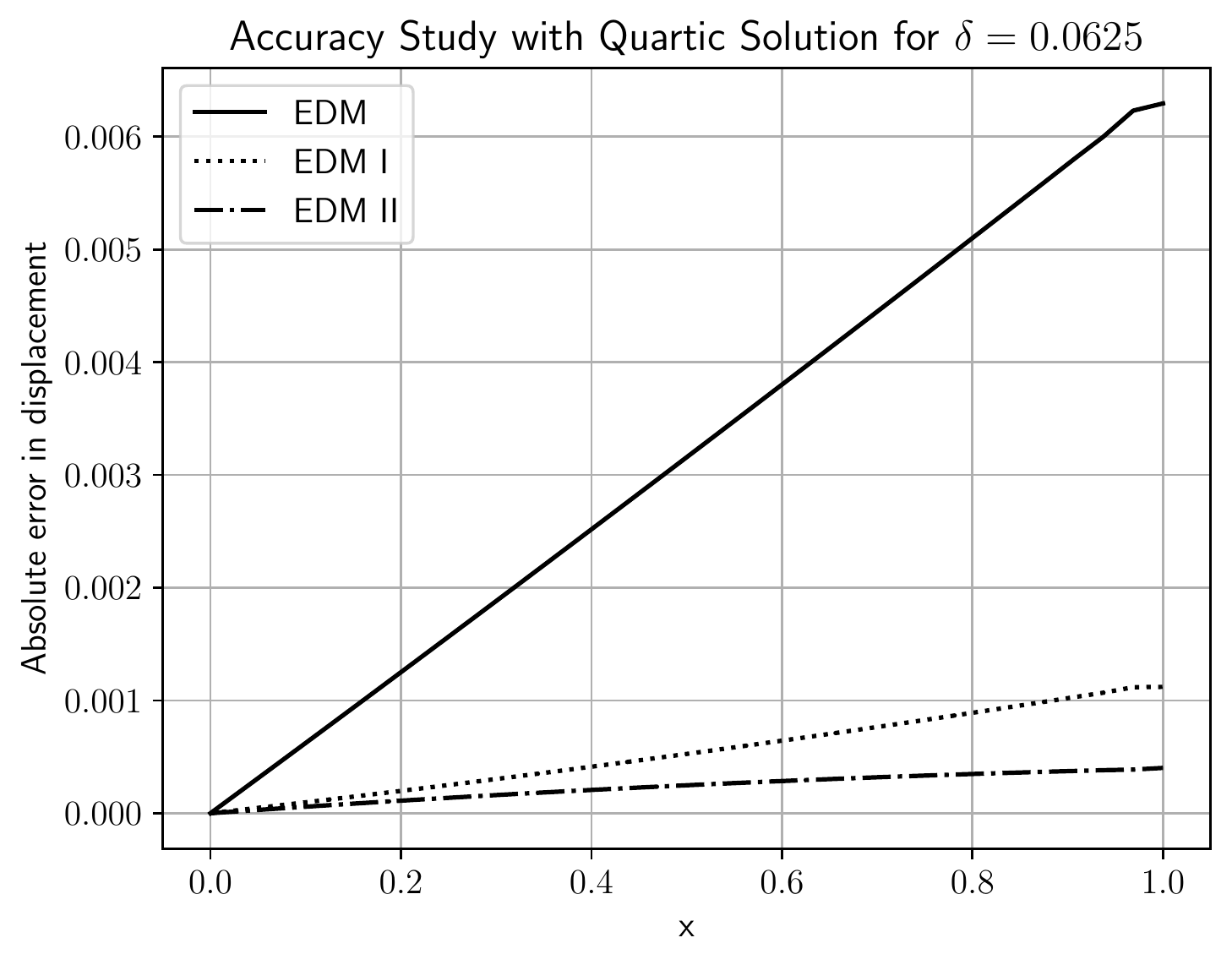}
    \caption{Comparison of the absolute error in displacement for EDM without correction, EDM I, and EDM II for $\delta=0.0625$.}
    \label{fig:EDMcomparison}
\end{figure}

\edit{In the previous numerical examples, we considered the extended domain method without any correction factor. In this example, we propose to study the effect of the analytical correction factor $\overline{k}(x)/\kappa$, as given in~\eqref{eq:EDM-correction-left} and~\eqref{eq:EDM-correction-right} and shown in Figure~\ref{fig:correctionfactorEDM} for $\delta=0.125$, and the numerical correction factor $\overline{k}(x_1)/\kappa = \overline{k}(x_{n-1})/\kappa = 8/7$ as explained by the result in~\eqref{eq:approx-secondderivative-87}. The correction methods with the analytical correction factor and the numerical factor shall be referred to here as EDM~I and EDM~II, respectively. We note that these correction factors are applied to the equations associated with node $x_1$ and node $x_{n-1}$ only. We consider the problem with the quartic solution as described in the previous section and plot in Figures~\ref{fig:EDMI} and~\ref{fig:EDMII} the errors in the displacement when using EDM~I and EDM~II for various values of the horizon $\delta$. We still observe some spurious effects in the solutions at $x_{n-1}$ for both methods, as in EDM, but these tend to diminish as the horizon goes to zero. We compare in Figure~\ref{fig:EDMcomparison} the absolute error in the displacement obtained with $\delta=0.0625$ for EDM without correction, EDM~I, and EDM~II. We observe that the maximum error in this case is reduced by a factor 5 for EDM~I and a factor 10 for EDM~II when compared to the EDM solution. Finally, the maximum relative errors are shown in Table~\ref{Tab:ErrorsEDMcomparison}. We observe that the rate of convergence for EDM without correction and EDM~I is about one, as expected. Surprisingly though, the convergence rate for EDM~II is approximately equal to two. We believe that, in this particular case, the solution benefits from favorable error cancellation as the relative error is even smaller than that obtained with LLEM and VHM, see Table~\ref{Tab:quartic} for $\delta=0.125$ and $\delta=0.0625$. In conclusion, we have demonstrated that it is possible to improve EDM using the analytical correction factor or the numerical correction factor. The latter seems to yield better results, but its value is directly dependent on the numerical scheme and should be derived for each type of differentiation stencil. On the other hand, the use of the analytical correction factor, while improving on EDM without correction, still leads to errors that are larger than VHM. For that reason, we will consider only VHM in the following numerical examples.} 

\begin{table}
\centering
\begin{tabular}{ccccc}
& & \multicolumn{3}{c}{Maximum relative error $\mathcal E_n$} \\
\midrule
$n$   & $\delta$  & EDM & EDM I & EDM II  \\
\midrule
16 & 0.125  
& \num{0.008889596332597536} & \num{0.0010471314156873602}  &  \num{0.0017700878124074493}\\
32 & 0.0625  
& \num{0.005114227123314417} & \num{0.0009174778126891541} & \num{0.0004478944065494997}  \\
64 & 0.03125  
& \num{0.0027285867770461512} & \num{0.0005588522036411084} & \num{0.00011264570888733981}\\
128 & 0.015625 
& \num{0.0014076939996958025} &  \num{0.0003046751610860376} & \num{2.8245458042428772e-05} \\
\bottomrule
\end{tabular}
\caption{Maximum relative error $\mathcal E_n$ for the quartic solution obtained with EDM without correction, EDM~I, and EDM~II for various values of the horizon $\delta$.}
\label{Tab:ErrorsEDMcomparison}
\end{table}

\subsection{m-convergence numerical study}

\begin{figure}
\centering
\includegraphics[width=0.9\columnwidth]{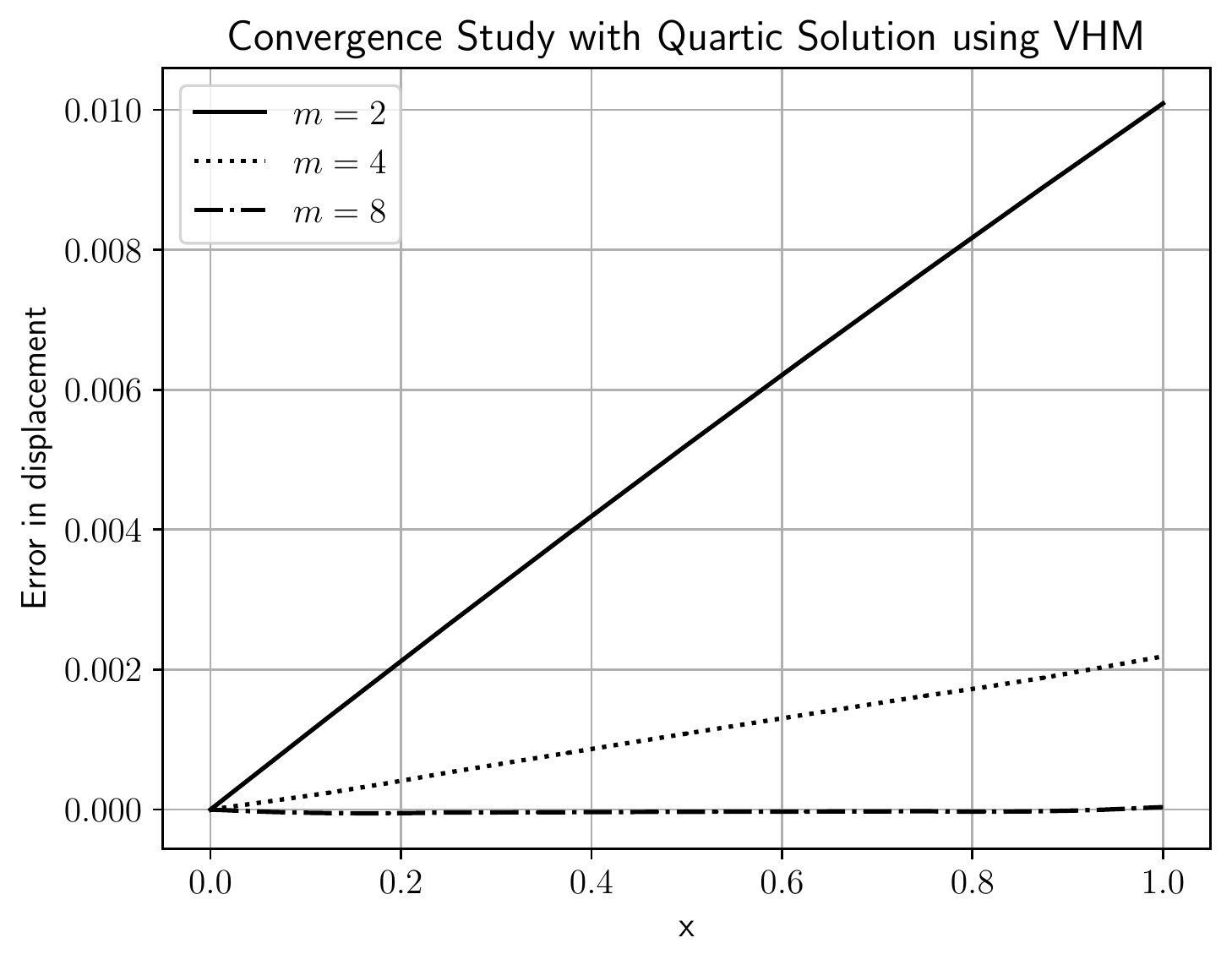}
\caption{Error in displacement for the quartic solution using the variable horizon method (VHM) for $\delta =1/4$ and various values of~$m$, with $h=\delta/m$.}
\label{Fig:VHM-Quartic-mconvergence-1}
\end{figure}

\begin{figure}
\centering
\includegraphics[width=0.9\columnwidth]{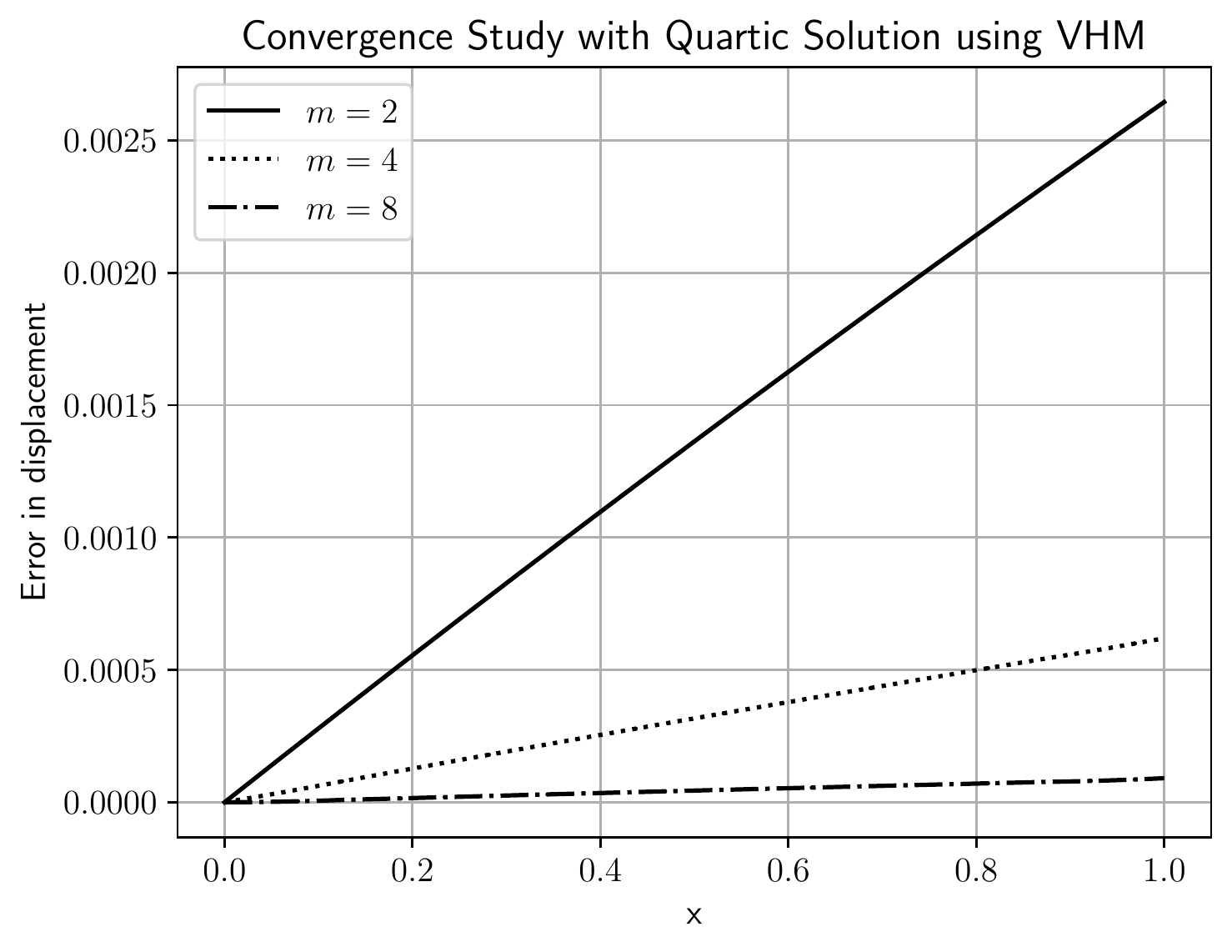}
\caption{Error in displacement for the quartic solution using the variable horizon method (VHM) for $\delta =1/8$ and various values of~$m$, with $h=\delta/m$.}
\label{Fig:VHM-Quartic-mconvergence-2}
\end{figure}

\corr{The objective of this numerical example is to show that the approximate solution by the variable horizon method converges to the exact solution of the peridynamic model. We choose for the manufactured solution the quartic function of the previous example, for which the corresponding loading term in the peridynamic problem is given by:
\[
f_b(x) = x^2 + \frac{\delta_v^2(x)}{12}.
\]
We consider here two values of the horizon, $\delta=1/4$ and $\delta=1/8$, and study the convergence in $h$, with $h=\delta/m$ and $m=2$, $4$, $8$. The maximum relative errors $\mathcal E_n$ and errors in the displacement field are shown in Table~\ref{Tab:VHM-Quartic-mconvergence-1} and Figure~\ref{Fig:VHM-Quartic-mconvergence-1} and in Table~\ref{Tab:VHM-Quartic-mconvergence-2} and in Figure~\ref{Fig:VHM-Quartic-mconvergence-2}, for $\delta=1/4$ and $\delta=1/8$, respectively. We clearly observe that the solutions converge to zero with respect to the discretization parameter $h$ at a rate of about two in both cases.
}

\begin{table}
    \centering
    \begin{tabular}{ccc}
    \toprule
    $m$   & $h$  & Maximum relative error $\mathcal E_n$  \\
    \midrule
2 & 0.125  & \num{0.008072916666666785} \\
4 & 0.0625  & \num{0.0017517300813050696} \\
8 & 0.03125  &  \num{0.00040901540792733805} \\\bottomrule
    \end{tabular}
\caption{Maximum relative error $\mathcal E_n$ for the quartic solution obtained with the variable horizon method (VHM) for $\delta=1/4$ and $h=\delta/m$, $m=2$, $4$, and $8$.}
\label{Tab:VHM-Quartic-mconvergence-1}
\end{table}

\begin{table}
    \centering
    \begin{tabular}{ccc}
    \toprule
    $m$   & $h$  & Maximum relative error $\mathcal E_n$  \\
    \midrule
2 & 0.0625 & \num{0.0021158854166701603} \\
4 & 0.03125 & \num{0.0004956586108461281} \\
8 & 0.015625 & \num{7.280434229048894e-05} \\\bottomrule
    \end{tabular}
\caption{Maximum relative error $\mathcal E_n$ for the quartic solution obtained with the variable horizon method (VHM) for $\delta=1/8$ and $h=\delta/m$.}
\label{Tab:VHM-Quartic-mconvergence-2}
\end{table}

\subsection{Manufactured solution with steep gradient}

\begin{figure}
\includegraphics[width=0.9\columnwidth]{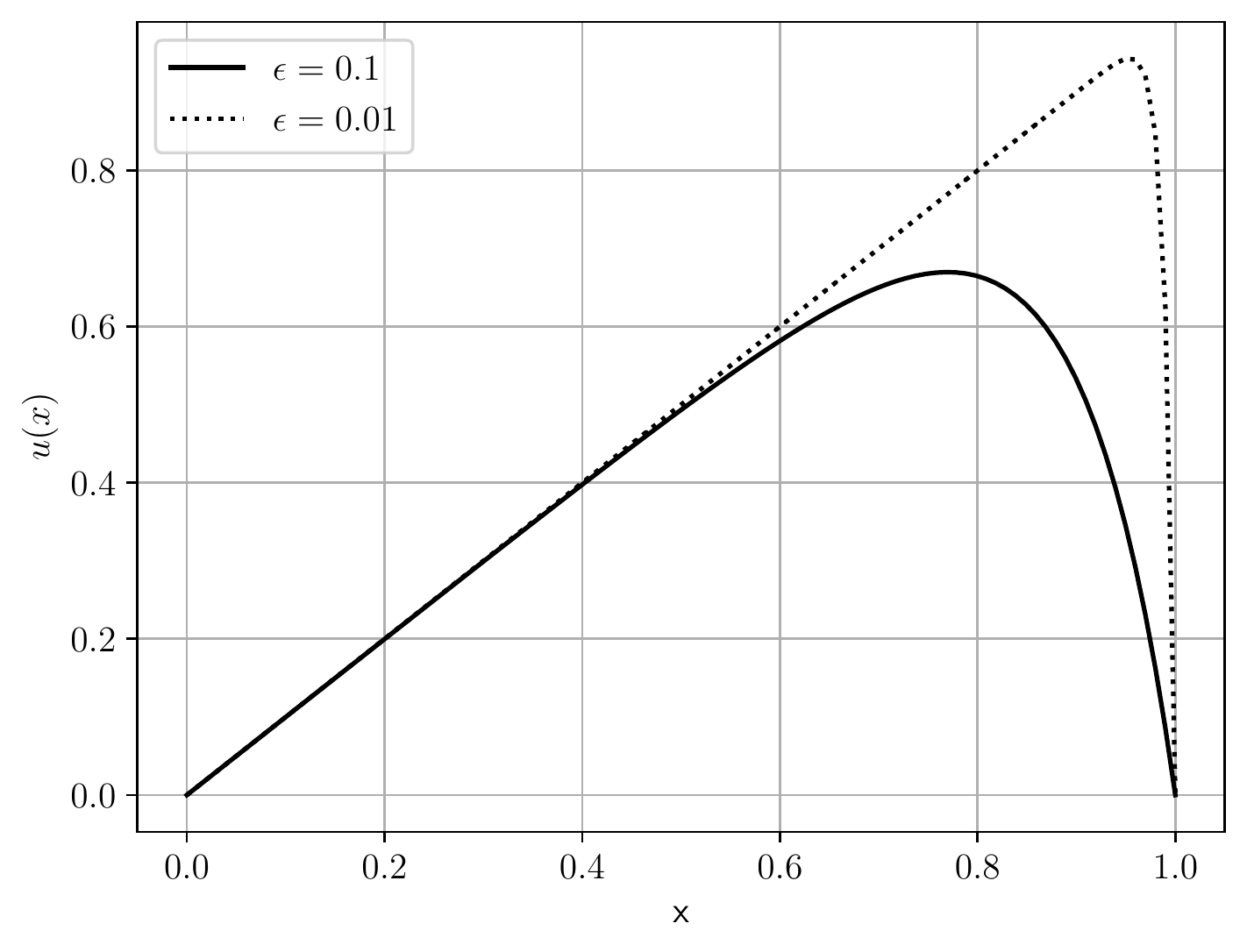}
\caption{Manufactured solution $\underline{u}(x)$ given in~\eqref{eq:exp-example} for $\epsilon=0.1$ and $\epsilon=0.01$.}
\label{Fig:displacement-exp}
\end{figure}

\corr{We consider in this example the manufactured solution to the classical linear elasticity problem:
\begin{equation}
\label{eq:exp-example}
\underline{u}(x) = x - \big(e^{-(1-x)/\epsilon}- e^{-1/\epsilon}\big) / \big(1-e^{-1/\epsilon}\big).
\end{equation}
The corresponding load $f_b(x)$ was calculated using the computer algebra system Maxima and the boundary data~$g$ for the Neumann boundary condition is given by:
\[
g = EA \underline{u}'(1) = 1 - \frac{1}{\epsilon(1 - e^{-1/\epsilon})}.
\] 
We observe that the magnitude of the first derivative $\underline{u}'$ at $x=1$ increases as $1/\epsilon$, as illustrated in Figure~\ref{Fig:displacement-exp} in the cases where $\epsilon=0.1$ and $\epsilon=0.01$.}

\corr{
Since the exact solution vanishes at $x=1$, we evaluate in this example the maximum absolute error $E_n$:
\[
E_n = \max_{i=1,\ldots,n} \left| \uelast(x_i) - u_i \right|,
\]
rather than the maximum relative error as before. 
These errors are reported in Tables~\ref{Tab:maxerror-exp-01} and~\ref{Tab:maxerror-exp-001} for $\epsilon=0.1$ and $\epsilon=0.01$, respectively. We observe that a large number of degrees of freedom, even more so when $\epsilon$ decreases, is necessary to attain small errors. However, it is noteworthy that the solutions still converge with a rate of two for VHM with respect to the horizon $\delta$. 
}

\begin{table}
    \centering
    \begin{tabular}{cccc}
    &  & \multicolumn{2}{c}{Maximum absolute error $E_n$} \\
    \toprule
    $n$  & $\delta$ & LLEM & VHM  \\\midrule
    32 & $\sfrac{1}{2^{16}}$ & \num{0.33800993851923344} & \num{0.41703479834078444}  \\
    64 & $\sfrac{1}{2^{32}}$ & \num{0.09258150807744349} & \num{0.11685617579259795}  \\
    128 & $\sfrac{1}{2^{64}}$  & \num{0.02421154115455596} & \num{0.03086476904103135}  \\
    256 & $\sfrac{1}{2^{128}}$  & \num{0.006178384378600433} & \num{0.00788181780929909} \\
    512 & $\sfrac{1}{2^{256}}$ & \num{0.0015539875133649187} & \num{0.0019653622588424384} \\\bottomrule    
    \end{tabular}
\caption{Maximum absolute error $E_n$ for the case with $\epsilon=0.1$ using the variable horizon method (VHM).}
\label{Tab:maxerror-exp-01}
\end{table}

\begin{table}
    \centering
    \begin{tabular}{cccc}
    & & \multicolumn{2}{c}{Maximum absolute error $E_n$} \\
    \toprule
    $n$  & $\delta$ & LLEM & VHM  \\\midrule
    512 & $\sfrac{1}{2^{256}}$ & \num{1.4149514799382532} & \num{1.7807642631472935}  \\
    1024 & $\sfrac{1}{2^{512}}$ & \num{0.37483362727391456} & \num{0.47926317088022624}  \\
    2048 & $\sfrac{1}{2^{1024}}$ & \num{0.09647827789907819}  & \num{0.12437326179253923}  \\
    4096 & $\sfrac{1}{2^{2048}}$ & \num{0.024474462722035573}   & \num{0.03168276910155437} \\
    8192 & $\sfrac{1}{2^{4096}}$ & \num{0.006163529946257666}  & \num{0.007995646147878224} \\\bottomrule    
    \end{tabular}
\caption{Maximum absolute error $E_n$ for the case with $\epsilon=0.01$ using the variable horizon method (VHM).}
\label{Tab:maxerror-exp-001}
\end{table}



\begin{figure}
\includegraphics[width=0.9\columnwidth]{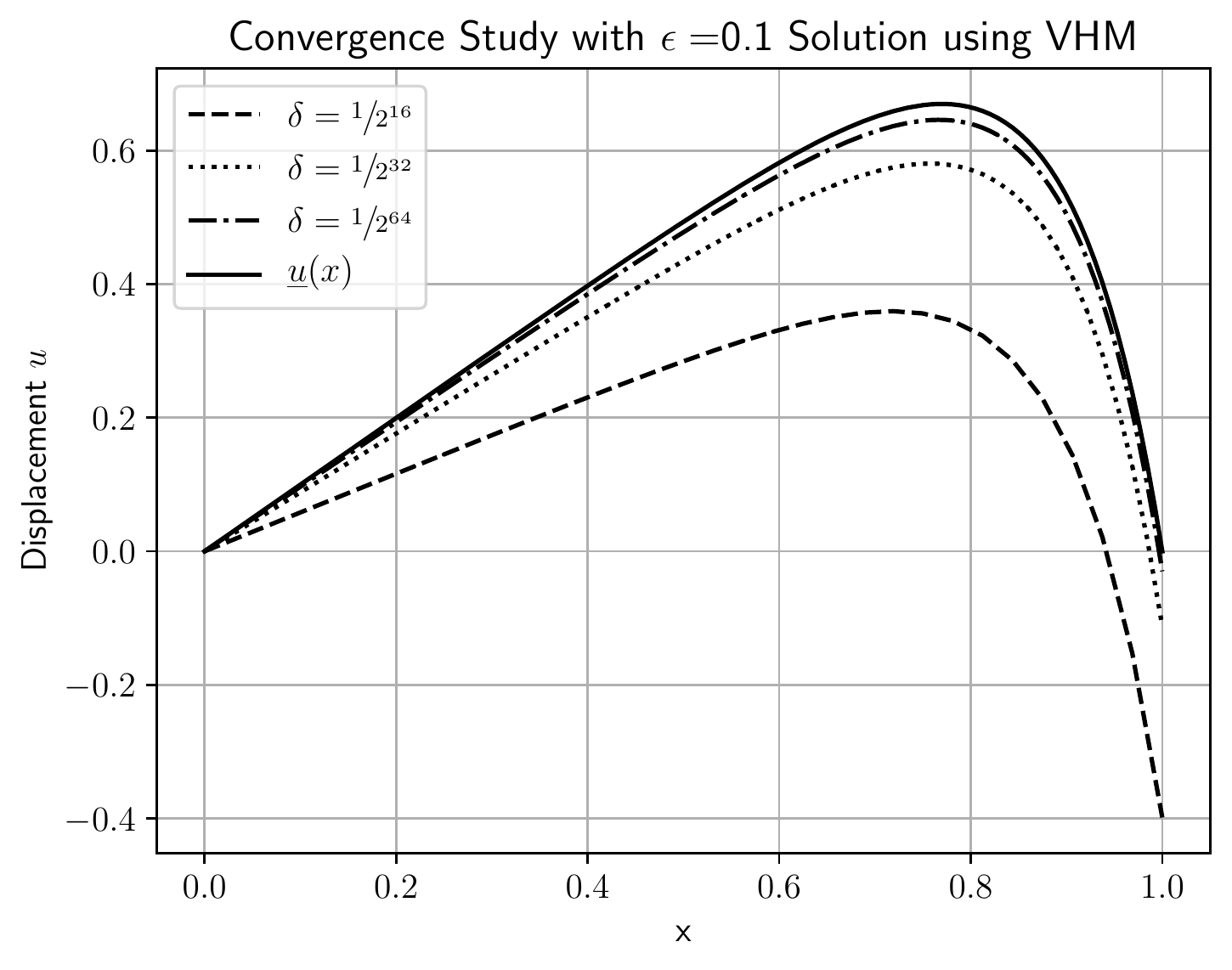}
\caption{Displacement for $\epsilon=0.1$ using the variable horizon method (VHM) with $\delta = 1/2^{16}$, $1/2^{32}$, and $1/2^{64}$.}
\label{Fig:displacement-exp-01}
\end{figure}

\begin{figure}
\includegraphics[width=0.9\columnwidth]{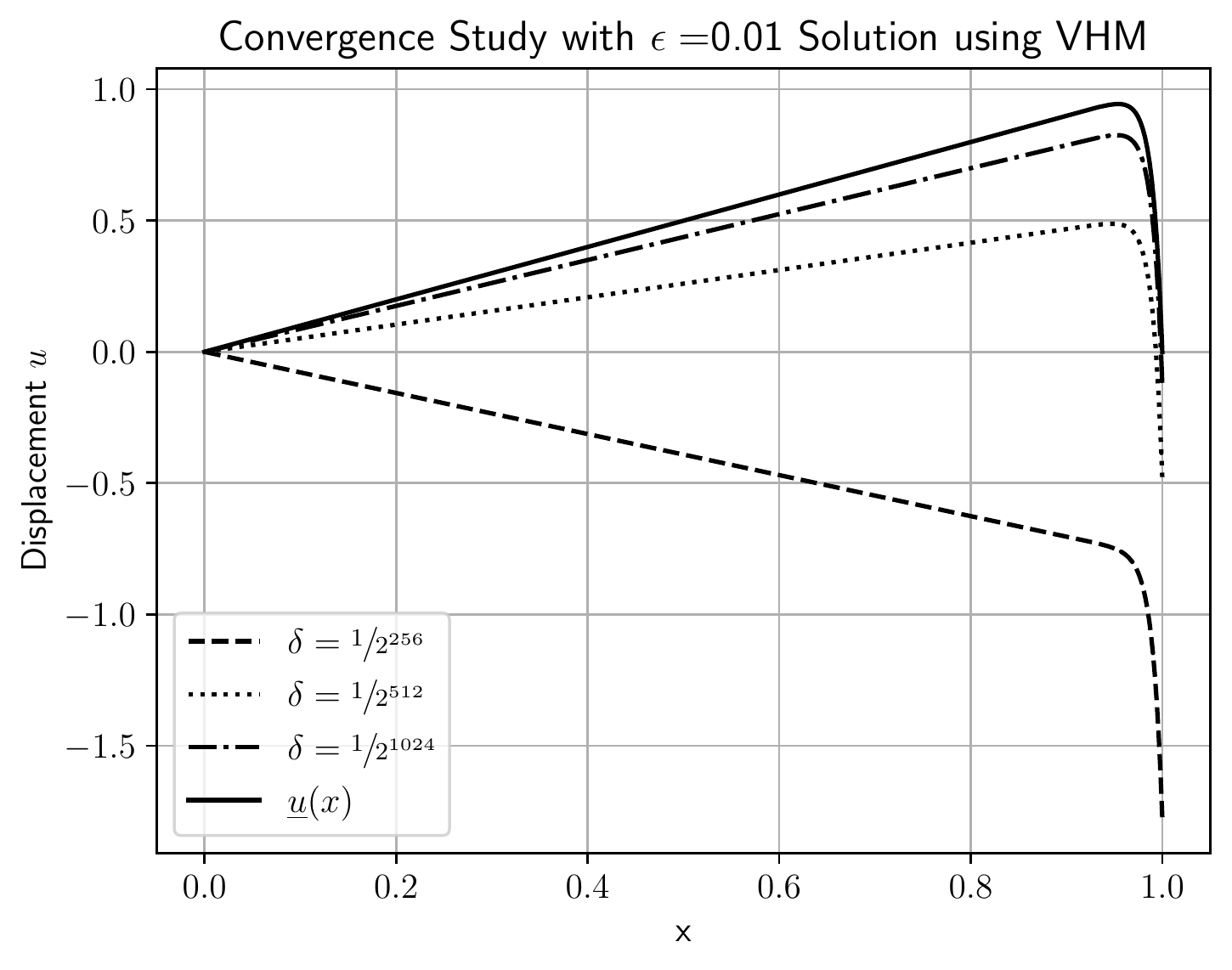}
\caption{Displacement for $\epsilon=0.01$ using the variable horizon method (VHM) with $\delta = 1/2^{256}$, $1/2^{512}$, and $1/2^{1024}$.}
\label{Fig:displacement-exp-001}
\end{figure}

\corr{
We show the displacement obtained with VHM for various values of $\delta$ for $\epsilon=0.1$ and $\epsilon=0.01$ in Figures~\ref{Fig:displacement-exp-01} and~\ref{Fig:displacement-exp-001}, respectively. It appears from these figures that the maximum error occurs mostly at the end point $x=1$ and that a large number of degrees of freedom needs to be employed in order to recover the boundary layer. It implies that the source of discretization error at $x=1$ severely pollutes the displacement field inside the domain.}

\section{Conclusion and future work}
\label{Sect:conclusion}

We have presented in this paper two methods, the so-called extended domain method and variable horizon method, to enforce boundary conditions within the non-local bond-based peridynamic model. \corr{The methods were developed based on the requirement that the solutions to the corresponding problems be compatible, as the horizon goes to zero, with the solution to the boundary-value problem derived from the classical linear elasticity model.} The formulation of both methods was first derived at the continuous level, independently of the choice of the discretization scheme, and later discretized, as an example, using the finite difference method. The performance of both methods was assessed on a simple one-dimensional linear elastic bar, fixed at one end and subjected to a traction at the other end, for various body force densities. The solutions of these manufactured problems were polynomial functions of degree one to four, which allowed us to study the advantages and disadvantages of each method.  

The extended domain method (EDM) without correction provided the correct solution only in the case \corr{of the problem with a linear solution}, for which the body force density is zero everywhere in the domain. In all other cases, we observed spurious artefacts in the displacement field near the boundaries. The reason for this behavior was explained by the fact that the extended domain method fails to correctly estimate the second derivative at the grid points closest to the boundaries, as it should in order to be compatible with the local model. \corr{We have seen that it was possible to improve the solutions obtained by EDM when using an analytical correction factor or a numerical correction factor. On the one hand, the analytical correction factor is derived at the continuous level and is valid for any type of discretization schemes. On the other hand, the numerical correction factor depends on the choice of the differentiation scheme. We observed on a numerical example that the numerical correction factor actually provided better results than the analytical correction factor. In view of this ambiguity, it is difficult to appreciate the superiority of one or the other correction approach.}

The variable horizon method (VHM) delivered, for all the numerical examples, consistent solutions, if not the same, with those obtained from the local linear elasticity model. In that sense, VHM should be considered a better approach than EDM when applying boundary conditions in non-local models. It is also worth noting here that VHM can be viewed as a coupling method between a local model and a non-local model (see e.g.~\cite{Zaccariotto-CMAME-2018}), where, in that particular case, the coupling interface happens to be at the boundary of the domain. 

VHM has shown its potential, through the preliminary one-dimensional numerical examples, for enforcing boundary conditions in the bond-based peridynamic model. \corr{In particular, we have theoretically and numerically shown that VHM is a method of order two in the horizon $\delta$, with respect to the linear elasticity model, as well as in the discretization parameter $h$, with respect to the finite difference method}. However, additional work should be carried out in order to confirm its performance in more general situations. For example, the method should be extended to two- and three-dimensional problems and assessed in the case of state-based peridynamic modeling. One could also consider other functions $\delta_v(x)$ than the piecewise linear function studied here and analyze their effects on the resulting solutions. Finally, the method should be assessed when other discretization methods, e.g.\ the Finite Element method, are used for its implementation.

\section*{Supplementary materials}
\noindent
The Python code (using numpy~\cite{oliphant2006guide,5725236}, scipy~\cite{2020SciPy-NMeth}, and matplotlib~\cite{Hunter:2007}) used to generate the numerical results is available on Github\footnote{\url{https://github.com/diehlpk/paperBCBBPD}} and on Zenodo~\cite{patrick_diehl_2019_3473692}.

\section*{Acknowledgements}
\noindent
\corr{Serge Prudhomme} is grateful for the support of this work by a Discovery Grant from the Natural Sciences and Engineering Research Council of Canada (Award \# RGPIN-2019-7154). \corr{Patrick Diehl} thanks the DTIC Contract FA8075-14-D-0002/0007 and the Center of Computation \& Technology at Louisiana State University for supporting this work.

\newpage
\bibliography{bibfile}

\end{document}